\documentstyle[12pt,epsf]{article} 
\input FEYNMAN
\def\gsim{\mathrel{\rlap {\raise.5ex\hbox{$ > $}}
{\lower.5ex\hbox{$\sim$}}}}
\def\lsim{\mathrel{\rlap {\raise.5ex\hbox{$ < $}}
{\lower.5ex\hbox{$\sim$}}}}

\newcommand{\be}{\begin{equation}}
\newcommand{\ee}{\end{equation}}
\newcommand{\bea}{\begin{eqnarray}}
\newcommand{\nn}{\nonumber}
\newcommand{\eea}{\end{eqnarray}}

\def\gappeq{\mathrel{\rlap {\raise.5ex\hbox{$>$}}
{\lower.5ex\hbox{$\sim$}}}}
 
\def\lappeq{\mathrel{\rlap{\raise.5ex\hbox{$<$}}
{\lower.5ex\hbox{$\sim$}}}}

\begin{document} 
\begin{titlepage} 

\begin{flushright} 
cond-mat/9912323
\end{flushright} 

\vspace{0.1in} 
\begin{centering} 

{\Large {\bf  Nodal Liquids in Extended t-J Models and  
Dynamical Supersymmetry}} \\
\vspace{0.2in} 
{\bf Nick E. Mavromatos } and {\bf Sarben Sarkar} \\
\vspace{0.1in} 
Department of Physics, King's College London \\
Strand, London WC2R 2LS, U.K. \\

\vspace{0.4in} 
{\bf Abstract}

\end{centering} 

{\small In the context of extended $t-J$ models, with intersite Coulomb
interactions of the form
$-V\sum\limits_{\left\langle {i,j} \right\rangle } {n_in_j}$, with $n_i$
denoting the  electron number
 operator at site $i$, nodal liquids are discussed. 
We use the spin-charge separation ansatz as applied
to the nodes of a d-wave superconducting gap.
Such a situation may be of relevance to 
the physics of high-temperature 
superconductivity. 
We point out the 
possibility of existence of certain points in the parameter space 
of the model characterized by dynamical supersymmetries 
between the spinon and holon degrees of freedom,  
which 
are quite different from the symmetries in conventional 
supersymmetric $t-J$ models. 
Such symmetries pertain to the continuum 
effective field theory of the nodal liquid, and 
one's hope is that the ancestor lattice model 
may differ from the continuum theory only by 
renormalization-group irrelevant operators in the infrared. 
We give plausible arguments that nodal liquids at 
such supersymmetric points are 
characterized 
by superconductivity of Kosterlitz-Thouless
type. 
The fact that 
quantum fluctuations around such points 
can be studied in a 
controlled way, probably makes such systems 
of special importance 
for an eventual non-perturbative understanding
of the complex phase diagram of the associated 
high-temperature superconducting 
materials.}

\vspace{0.90in}

\end{titlepage}

\section{Introduction}

The study of strongly correlated electron systems (SCES) is a major 
enterprise in modern condensed matter physics primarily due to high
temperature
(planar) superconductors, fractional Hall conductors,
and, more recently, in semiconductor quantum dots.
Owing to various non-Fermi liquid features of SCES, many believe that the
 low-energy excitations of 
these systems are influenced by the proximity of a critical Hamiltonian
in a generalized coupling-constant space. In this scenario,
known as spin-charge separation~\cite{anderson},
these excitations are spinons, holons and gauge fields.

Important paradigms for SCES are the conventional Hubbard model, 
or its $t-j$ extension, both of which have been conjectured
to describe the physics of high-temperature 
superconducting doped antiferromagnets. 
Numerical simulations of such models~\cite{dagotto}, 
in the presence of
very-low doping, have provided evidence for electron substructure
(spin-charge
separation) in such systems.

In ref. \cite{[5]},  
an extension of the spin-charge separation
ansatz, allowing
for a particle-hole symmetric formulation away from half-filling,
was introduced by writing:
\be
\chi _{\alpha \beta }\equiv \left( {\matrix{{\psi _1}&{\psi _2}\cr
{-\psi _2^\dagger}&{\psi _1^\dagger}\cr
}} \right)_i\left( {\matrix{{z_1}&{-\bar z_2}\cr
{z_2}&{\bar z_1}\cr
}} \right)_i\,,\label{1.5}
\ee
where
the fields $z_{\alpha,i}$ obey canonical {\it bosonic}
commutation relations, and are associated with the
{\it spin} degrees of freedom (`spinons'), whilst the fields $\psi$
are Grassmann variables, which obey Fermi 
statistics, and are associated with the electric charge degrees of freedom
(`holons').
There is a hidden non-abelian gauge symmetry~
$SU(2) \otimes U_S(1)$ in the ansatz,
which becomes a dynamical symmetry
of the pertinent planar Hubbard model, studied in ref. \cite{[5]}. 

The ansatz (\ref{1.5}) is different from that of refs.~\cite{[17],fisher},
where the holons are represented as charged bosons, and the spinons as
fermions. That framework, unlike ours, is not a convenient starting point
for making predictions such as the behaviour of the
system under the influence of strong external fields.
As argued in \cite{fmijmpb}, 
a strong magnetic field induces the opening of a second
 superconducting gap at the nodes of the $d$-wave gap, in agreement
with recent experimental findings on the behaviour of the thermal
conductivity
of high-temperature cuprates 
under the influence of strong external magnetic fields~\cite{ong}.

In \cite{[5]} a single-band 
Hubbard model was used. Such a model 
should not be regarded as merely phenomenological
for cuprate superconductors since it can be deduced
from chemically irealistic multiband models involving both Cu and O 
orbitals and it has extra
nearest-neighbour interactions of the form~\cite{[1]}:
\be
H_{int} = -V \sum_{<ij>}
n_i n_j \qquad n_i \equiv \sum_{\alpha=1}^{2}
c_{\alpha,i}^\dagger c_{\alpha,i}\,,\label{1.7}
\label{extraint}
\ee
as well as longer finite-range hoppings.

What we shall argue below is that the presence of 
interactions of the form (\ref{extraint}) is crucial 
for the appearence of supersymmetric points
in the parameter space of the spin-charge 
separated model. Such points 
occur for particular doping concentrations.
As we shall discuss, 
this supersymmetry 
is a {\it dynamical symmetry}
of the spin-charge separation, 
and occurs between 
the spinon and holon degrees of freedom 
of the ansatz (\ref{1.5}). 
Its appearance may indicate the onset of 
unconventional 
superconductivity of the Kosterlitz-Thouless (KT) type~\cite{[2],[3]} 
in the liquid of excitations about the nodes of the 
d-wave superconducting gap
(``nodal liquid''), to which we restrict our attention
for the purposes of this work. 

It should be stressed that 
the supersymmetry characterizes the 
continuum {\it relativistic} effective (gauge) 
field theory of the nodal 
liquid.
The progenitor lattice model is of course
{\it not supersymmetric} in general. What one hopes, however, 
is that at such supersymmetric points the 
universality 
class of the continuum low-energy 
theory is the {\it same} 
as that of the lattice model, in the sense that the latter
differs from the continuum effective theory only by 
renormalization-group {\it irrelevant} operators (in the infrared). 
This remains to be checked by detailed
studies, which do not constitute the topic of this article. 

In general, supersymmetry 
provides a much more 
controlled way for 
dealing with quantum fluctuations about
the ground state of a field-theoretic system
than a non-supersymmetric theory~\cite{[14]}. 
In this sense, one hopes that by working in such 
supersymmetric points in the parameter space of the 
nodal liquid she/he might obtain some exact results 
about the phase structure, which might be useful 
for a non-perturbative understanding of the 
complex phase diagrams that characterize the 
physics of the (superconducitng) doped antiferromagnets. 
As we shall discuss below, to obtain supersymmetric points one needs 
to make specific asumptions about the regime of the 
parameters of the model; from an energetics point of view, 
such assumptions are retrospectively 
justified 
by the
fact that supersymmetric ground states 
are characterized by {\it zero energy}~\cite{[14]}, and 
hence are acceptable ground 
states from this point 
of view. 

Significant progress towards a non-perturbative 
understanding of Non-Abelian gauge field theories, 
in four space-time dimensions, 
based on supersymmetry has been made 
by Seiberg and Witten~\cite{[15]}.
The fact that the spin-charge separation 
ansatz (\ref{1.5}) of the doped antiferromagnet is known to be characterized
by such non-Abelian gauge  structure is an encouraging sign. 
However, it should be noted that in the case of ref. \cite{[15]}
extended supersymmetries were necessary for yielding exact results.
As we shall discuss below, under special 
conditions for doped 
antiferromagnets,
the supersymmetric points are characterized by $N=1$ 
three-dimensional supersymmetries.
Under certain circumstances the supersymmetry may be elevated
to $N=2$~~\cite{[8]}, for which it is possible to
obtain 
some exact results concerning the phase structure~\cite{[16]}. 
In the present state of the udnerstanding 
of SCES it is a pressing need to have 
relevant models for which we can extract 
non-trivial exact information.
However, for a realistic 
condensed-matter system such as a high-temperature
superconductor, 
even the $N=1$ supersymmetry
of the supersymmetric points is expected to be broken 
at finite temperatures or under the influence of external 
elctromagnetic fields.  
Nevertheless, one may hope that by viewing 
the case of broken supersymmetry as a 
perturbation about 
the supersymmetric point,
valuable non-perturbative information may still be obtained.
As we shall see, a possible example of this may be 
the above-mentioned KT superconducting properties~\cite{[2]} 
that characterize such points. 

The structure of the article is as follows: 
In section 2 we describe briefly 
the statistical model
which gives rise to the continuum relativistic 
effective (2+1)-dimensional field theory of the nodal liquid.
In section 3 we discuss the properties and
(non-abelian gauge) symmetries  of the spin-charge separation
ansatz that characterizes the model. 
In the next section we discuss the intersite Coulomb 
interactions, which are of crucial importance for the 
existence of supersymmetric points. In section 5 we 
state the conditions for N=1 supersymmetry at such points,
and describe briefly their importance for 
yielding superconductivity of Kosterlitz-Thouless type. 
We conclude in section 6 with some prospects for future 
work. Technical aspects of our work, which may 
help the non-expert reader to follow the arguments 
presented in the text in a mathematically detailed 
way, are given in two Appendices. 

\section{The Model and its Parameters}

In reference \cite{[1]} 
it was argued that BCS-like scenarios for high $T_c$ superconductivity based
on
 extended $t-J$ models yield reasonable predictions for the critical
temperature $T_c^{\max }$ at optimum doping . 
There it was argued that a pivotal role was played by 
next-to-nearest neighbour and third neighbour hoppings, ${t'}$ 
and ${t''}$ respectively. In particular the combination 
$t_-\equiv t'-2t''$ determines the shape of the Fermi surface 
and the nature of the saddle points and the associated $T_c^{\max }$.

Our aim is to use the extended $t-J$ model studied  
in \cite{[1]} in order to discuss the appearance of relativistic 
charge liquids at the nodes of the associated d-wave superconducting gap. We
will argue that the nodes characterize
the model in a certain range of parameters. 
We will demonstrate that at a certain 
regime of the parameters and doping 
concentration the nodal liquid effective 
field theory of spin-charge separation 
exhibits supersymmetry. This supersymmetry is dynamical and should not be
confused with the non-dynamical symmetry
under a graded supersymmetry algebra that characterises the spectrum of
doped antiferromagnets at two special points of
the parameter space \cite{[4]}. We shall also 
discuss unconventional mechanisms for superconductivity
in the nodal liquid similar to 
the ones proposed in \cite{[2],[3]}.

To start with let us describe briefly 
the extended $t-J$ model used in Ref. \cite{[1]}. 
The Hamiltonian is given by:
\be
H=P\left( {H_{hop}+H_J+H_V} \right)P + PH_\mu P\,,\label{1.1}
\ee
where: 

(a)
\be 
H_{hop}=-\sum\limits_{\left\langle {ij} \right\rangle } {t_{ij}c_{i\alpha
}^+
c_{j\alpha }-\sum\limits_{\left[ {ij} \right]} {t'_{ij}}}c_{i\alpha
}^+c_{j\alpha }-\sum\limits_{\left\{ {ij} \right\}} 
{t''_{ij}}c_{i\alpha }^+c_{j\alpha }\,,\label{2.2}\ee
and $\left\langle {\dots } \right\rangle $ 
denotes nearest neighbour (NN) sites, $\left[ {} \dots 
\right]$ next-to-nearest neighbour (NNN), and
$\left\{ {} \right\}$ third nearest neighbour.
Here repeated spin (or "colour") indices 
are summed over. The Latin indices $i,j$ 
denote lattice sites and the Greek indices $\alpha =1,2$
are spin components. 

(b) \be H_J=J\sum\limits_{\left\langle {ij} \right\rangle } 
\left(T_{i,\alpha \beta } T_{j,\beta \alpha } 
-\frac{1}{4}n_i n_j\right) +
J'\sum\limits_{\left[ {ij} \right]} 
{T_{i,\alpha \beta }}T_{j,\beta \alpha }\,,\label{2.3}\ee
with $n_i=\sum\limits_{\alpha =1}^2 {c_{i\alpha }^+
c_{i\alpha }}$, and 
$T_{i,\alpha \beta }=c_{i\alpha }^+c_{i\beta }$.
The quantities $J,J'$ 
denote the couplings of the 
appropriate Heisenberg antiferromagnetic interactions.  
We shall be interested~\cite{[3]} in the regime where 
$J'<<J$.

(c) \be H_\mu =\mu \sum\limits_i {c_{i\alpha }^+c_{i\alpha }}\,,
\label{2.4}
\ee and $\mu$ is the chemical potential.

(d) \be H_V=-V\sum\limits_{\left\langle {ij} \right\rangle } {n_i}n_j\,,
\label{2.5a}
\ee
This is an effective static NN interaction which, in the 
bare $t-J$ model, is induced by the exchange term, 
because of the extra magnetic bond 
in the system 
when two polarons are on neighbouring sites~\cite{[1]}. 
Notice that this term, when combined with the Coulomb interaction
terms in $H_J$, yields in the effective action 
a total inter-site Coulomb interaction term with coupling
\be
    V_{total} = V + 0.25~\,J 
\label{totalv}
\ee
In
ref. \cite{[1]} the strength of the interaction (\ref{2.5a}) is taken to be:
\be V\approx 0.585\,J\,,\label{2.6a} \ee
This is related to the regime of the parameters used in \cite{[1]},
for which the NN hoping element satisfies $t << J$. 
In fact, for the effective $t-j-V$ model of \cite{[1]},
viewed as an appropriate reduction of a  single-band Hubbard 
model, one has the relation:  
\be
J=\frac{4t^2}{U_{eff}+ V'} + J_{SB}
\label{param}
\ee
where $U_{eff}$ is an effective Hubbard interaction, and $J_{SB}$ is 
a ferromagnetic exchange Heisenberg energy for the single-band model.
We have $\left| {V'} \right|\ne \left| V \right|$ in general, unlike the
case of the standard Hubbard model with a supplementary intersite Coulomb
interaction. 
 However, one may consider more general models,
in which the above restriction 
is not imposed, and $V$ is viewed as an independent 
parameter of the effective theory, e.g. 
\be V\approx b~\,J\,,\label{2.6ab} \ee
where $b$ is a constant to be determined phenomenologically. 
Such a situation may arise, for instance, in 
effective models where one considers repulsive on-site Coulomb
interactions~\cite{[1]} 
(e.g. between holes and/or electrons) {\it in addition} to 
the (electron-hole) attractions (\ref{2.5a}).
As we shall discuss below, 
such more general cases turn out to be useful 
for the existence of 
supersymmetric points in the parameter space of the model.

(e) The operator $P$ is a projector operator, expressing the
absence of double occupancy at a site.

We define the doping parameter $0 < \delta  < 1$ by 
\be
{\left\langle n_i \right\rangle} =1-\delta \,,\label{2.7}\ee

$d$-wave pairing, which seems to have been confirmed experimentally for
high-$T_c$ cuprates, was assumed in 
\cite{[1]}. 
A d-wave gap is represented by an order parameter of the form
\be
\Delta \left( {\vec k} \right)
=\Delta _0\left( {\cos \,k_xa-\cos \,k_ya} \right) \,,\label{2.8b}
\ee
where $a$ is the lattice spacing.
The relevant Fermi surface is characterised by the following four 
nodes where the gap vanishes:
\be
\left( {\pm {\pi  \over {2a}},\pm {\pi  \over {2a}}} \right)\,,
\label{2.9b}\ee
We now consider the generalized dispersion relation \cite{[5],[6]} 
for the quasiparticles in the superconducting state:
\be
E\left( {\vec k} \right)=\sqrt {\left( {\varepsilon \left( {\vec k}
\right)-\mu } \right)^2+\Delta ^2\left( {\vec k} \right)}\,,\label{2.10b}
\ee
In the vicinity of the nodes it is 
reasonable~\cite{[5],[6]} 
to assume that $\mu \approx 0$ or equivalently  we may
linearize about $\mu$, 
i.e. write $\varepsilon \left( {\vec k} \right)-\mu \approx v_D\left| {\vec
q} \right|$~\cite{[2]} where
$v_D$ is the 
effective velocity at the node and $q$ is the wave-vector with 
respect to the nodal point.

\section{Non-Abelian spin-charge separation in the t-J model}

As already mentioned in the introduction, 
it was {\it proposed} in ref. \cite{[5]} 
that for the large-U limit of the {\it doped} 
Hubbard model the following {\it `particle-hole' symmetric 
spin-charge separation} 
ansatz occurs at {\it each site} $i$: 

\be\chi _{\alpha\beta,i} = 
\psi _{\alpha\gamma,i}z_{\gamma\beta,i} \equiv \left(
\begin{array}{cc}
c_1 \qquad c_2 \\
c_2^\dagger \qquad -c_1^\dagger \end{array} 
\right)_i = 
\left(\begin{array}{cc} 
\psi_1 \qquad \psi_2 \\
-\psi_2^\dagger \qquad \psi_1^\dagger\end{array} 
\right)_i~\left(\begin{array}{cc} z_1 \qquad -{\overline z}_2 \\
z_2 \qquad {\overline z}_1 \end{array} \right)_i \label{3.1}
\ee
where the fields $z_{\alpha,i}$ obey canonical {\it bosonic} 
commutation relations, and are associated with the
{\it spin} degrees of freedom (`spinons'),  
whilst the fields 
$\psi _{a,i},~a=1,2$ have {\it fermionic}
statistics, and are assumed to {\it create} 
{\it holes} at the site $i$ with spin index $\alpha$ (`holons'). 
The ansatz (\ref{3.1}) 
has spin-electric-charge separation, since only the 
fields $\psi$ carry {\it electric} charge. Generalization to the non-abelian
model allows for inter-sublattice hopping of holes which is observed
experimentally.

It is worth noticing that the anti-commutation relations 
for the electron fields $c_\alpha$,$c_\beta^\dagger$,
do not quite follow from the ansatz (\ref{3.1}). 
Indeed, assuming the canonical 
(anti-)commutation relations for the $z$ ($\psi$) fields, 
one obtains from the ansatz (\ref{3.1}) 
\bea
\{ c_{1,i}, c_{2,j} \} & \sim & 2 \psi_{1,i}\psi _{2,i}\delta_{ij} \nn \\
\{ c_{1,i}^\dagger, c_{2,j}^\dagger \} &\sim & 2 \psi_{2,i}^\dagger
\psi _{1,i}^\dagger \delta_{ij} \nn \\
\{ c_{1,i}, c_{2,j}^\dagger \} & \sim& \{ c_{2,i}, c_{1,j}^\dagger \}
\sim 0 \nn \\
\{ c_{\alpha,i}, c_{\alpha,j}^\dagger \} & \sim &  
\delta _{ij}\sum_{\beta =1,2} [z_{i,\beta} {\overline z}_{i,\beta} + 
\psi_{\beta,i}\psi _{\beta,i}^\dagger],~~\alpha=1,2
\qquad {\rm no~sum~over~i,j}\label{3.2}\eea
To ensure the {\it canonical} anticommutation 
relations for the $c$ operators
we must therefore {\it impose} 
at each lattice site the 
(slave-fermion) constraints
\bea &~&\psi_{1,i}\psi_{2,i} 
= \psi^\dagger _{2,i}\psi^\dagger_{1,i} = 0,\nn \\
&~&\sum_{\beta =1,2} [z_{i,\beta} {\overline z}_{i,\beta} + 
\psi_{\beta,i}\psi _{\beta,i}^\dagger] = 1
\label{3.3}
\eea
Such relations are understood to be satisfied when the 
holon and spinon operators act on {\it physical} states.
Both of these relations are valid in the large-$U$ limit 
of the Hubbard model and encode the non-trivial physics 
of constraints behind the spin-charge separation ansatz 
(\ref{3.1}). They express the 
constraint of {\it at most one electron or hole per site},
which characterizes the large-$U$ Hubbard models we are considering 
here. 

There is a local phase (gauge) non-abelian symmetry hidden in the 
ansatz (\ref{3.1})~\cite{[5]}
$G=SU(2)\times U_S(1)$, where $SU(2)$ stems from 
the spin degrees of freedom, $U_S(1)$ is a 
statistics changing group, which 
is exclusive to two spatial dimensions and is responsible 
for transforming bosons into fermions and vice versa.
As remarked in \cite{[5]}, 
the $U_S(1)$ effective interaction is responsble for the 
equivalence between  the slave-fermion ansatz  
(i.e. where the holons are viewed as 
charged bosons and the spinons as electrically 
neutral fermions~\cite{[17]})
and the slave boson ansatz (i.e. where the holons are viewed as 
charged fermions and the spinons as neutral bosons~\cite{[18],[5]}). 
This is analogous (but not identical) to the bosonization 
approach of \cite{[19]} for anyon systems. 

The application of the ansatz (\ref{3.1}) 
to the Hubbard (or t-j models) 
necessitates a `particle-hole' symmetric formulation of the 
Hamiltonian (\ref{1.1}), 
which as shown in \cite{[5]}, 
is expressible
in terms of the operators $\chi$. 
In this way, for instance, the NN Heisenberg interactions terms 
become:
\be 
H_J = -\frac{1}{8}\sum_{<ij>} {\rm Tr}\left[ \chi_i \chi^\dagger_j 
\chi_j \chi^\dagger_i \right]
\label{heischi}\ee 
By making and appropriate 
Hubbard-Stratonovich 
transformation on $H_J$ with 
Hubbard-Stratonovich fields $\Delta_{ij}$, 
we obtain 
the effective spin-charge separated action
for the doped-antiferromagnetic model of \cite{[5]}:
\bea
&~& H_{HF}=\sum_{<ij>} \left(tr\left[(8/J)\Delta^\dagger_{ij}\Delta_{ji}
+ \left| A_1 \right| (-t_{ij}(1 +
\sigma_3)+\Delta_{ij})\psi_j V_{ji}U_{ji}\psi_i^\dagger\right] + \right.\nn
\\
&~& \left. tr\left[ K{\overline z}_iV_{ij}U_{ij}z_j\right] + h.c. \right)
+ \dots \,,
\label{1.6}
\eea
with the $\dots$ denoting 
chemical potential terms and NNN hopping terms (the latter are essential 
for the model of \cite{[1]}; we shall discuss their effects below).

This form of the action, 
describes low-energy excitations 
about the Fermi surface of the theory. 
The field $\Delta_{ij}$ is matrix valued in `colour' 
space; generically it may be expanded in components in a canonical 
basis of $2 \times 2$ matrices, $ \{1,\sigma^a \},~a=1,2,3$,
as follows:
\be
\left(\Delta_{ij}\right)_{\alpha\beta} = A_0 \delta_{\alpha\beta}
+ A_a \left(\sigma^a \right)_{\alpha\beta}
\label{deltadef}
\ee 
where Greek indices denote $2\times 2$ `colour' indices.

The quantities
$V_{ij}$ and $U_{ij}$ denote lattice link variables associated with 
elements of the  $SU(2)$ and $U_S(1)$ groups respectively.
They are associated~\cite{[5]} 
with phases 
of vacuum expectation values 
of bilinears $<~{\overline z}_i z_j~>$ 
and/or $<~\psi^\dagger _i \left( -t_{ij}(1 + \sigma _3) + \Delta _{ij}
\right)\psi_j~>$. It is understood that, 
by integrating out in a path integral over $z$ and $\psi$ variables,
fluctuations are incorporated, which go beyond a Hartree-Fock treatment.

The quantity $\left|A_1 \right|$ is the amplitude of the bilinear 
$<~{\overline z}_i z_j~>$ assumed frozen~\cite{[5]}. 
By an appropriate normalization 
of the respective field variables, one may set $\left|A_1 \right| =1$,
without loss of generality. 
In this normalization, one may then 
parametrize the quantity $K$, which is the amplitude of 
the appropriate
fermionic bilinears, as~\cite{[5],[3]}:
\be
K \equiv \left(J \left| \Delta _z \right|^2  (1-\delta)^2 \right)^{1/2}~;
\qquad   
1 - \delta = <\sum_{\alpha =1}^{2}~\psi_\alpha \psi^\dagger_\alpha >~, 
\label{defK}
\ee
with $\delta $ the doping concentration in the sample.  
The quantity  
$|\Delta _z |$ is considered as an arbitrary 
parameter of our effective theory, of dimensions $[energy]^{1/2}$, 
whose magnitude  
is to be fixed 
by phenomenological or other considerations (see below). 
To a first approximation we assume that $\Delta _z $
is doping independent~\footnote{However, from its 
definition,  as a $< \dots >$ 
of a quantum model with complicated $\delta $ dependences in its 
couplings, the quantity $\Delta _z$ may indeed exhibit 
a doping dependence. For some consequences of this 
we refer the reader to the discussion in section 6, below.}.   
The dependence on $J $ and $\delta$ in (\ref{defK}) 
is dictated~\cite{[3]} by the
correspondence with the conventional antiferromagnetic
$CP^1$ $\sigma$-model in the limit $\delta \rightarrow 0$. 

The model of ref. \cite{[1]} differs from that of \cite{[5]}
in the existence of 
NNN hopping $t'$ and 
triple neighbour hopping $t''$, which were ignored in the 
analysis of \cite{[5]}.  
For the purposes of this work, 
which focuses on the low-energy (infrared) 
properties
of the continuum field theory of (\ref{1.6}), 
this can be taken into 
acount 
by assuming that 
\be
|t_{ij}| = t'_{+} \equiv  t + 2t_+, 
\qquad t_+ \equiv t' + 2t''\label{2.6b}
\ee
in the notation of \cite{[1]}. 
The relation stems from the observation that 
in the continuum low-energy 
field-theory limit such 
$NNN$ and triple hopping terms
can be Taylor expanded (in derivatives). 
It is the terms linear in derivatives that 
yield the shift (\ref{2.6b}) 
of the NN neighbour hopping element $t$.  
Higher derivatives terms, 
of the form $\partial _x \partial _y$ are suppressed 
in the low-energy (infrared) limit. 

It is important to note that the model of \cite{[5]},
as well as its extension (\ref{1.6}),  
in contrast to that discussed in \cite{[2]},  
involves only a {\it single lattice} structure, with nearest neighbour
hopping ($<ij>$) being taken into account, $t_{ij}$. 
The antiferromagnetic nature is then viewed as a 
property of a `colour' degree of freedom,
expressed via the non-abelian gauge structure of the 
spin-charge separation ansatz (\ref{3.1}). 
As we shall discuss later, 
this is very important in yielding the correct number of fermionic 
(holons $\Psi$) 
degrees of freedom in the continuum low-energy field theory to match 
the bosonic degrees of freedom (spinons $z$) at the supersymmetric point. 

\section{The Effective Low-Energy Gauge Theory}

\subsection{Nambu-Dirac Spinor Representation of nodal Holons}

It is instructive to discuss in some detail 
the derivation of 
a conventional lattice gauge theory form of the action 
(\ref{1.6}).
One first 
shifts the 
$\Delta_{ij}$ field: $\Delta_{ij} \rightarrow \Delta '_{ij} =\Delta_{ij} +
t_{ij}\sigma_3$, 
and then assumes that 
the fluctuations of the $\Delta '_{ij}$ 
field are frozen in such a way that only the $<A'_0>$ component 
is non trivial in the corresponding 
expansion in terms of the Pauli matrices
(\ref{deltadef}).
This is a variational ansatz that  
can be justified in the regime 
of the parameters of the statistical model
$J >> t_+' $, in which case the dominant 
$\Delta_{ij}$ configurations (in the path integral) 
may be taken to be of order $J$, and thus any effect of the $\sigma_3$ 
colour structure in the action (\ref{1.6}) is safely negligible.
As we shall discuss in what follows, the elimination of the $\sigma_3$ terms
from the action (\ref{1.6}) results in canonical Dirac kinetic terms
for the fermionic parts of the nodal liquid effective (low-energy) action. 

However, 
in view of (\ref{2.6b}), in the model of \cite{[1]}, such an assumption is
not 
valid, given that the renormalized hopping parameter, due to NNN and
triple neighbour hoppings, is of similar order as $J$. Nevertheless, 
for our generic purposes in this work we shall  
work in a model where $J >> t_+'$. Alternatively, we can assume 
that the effects of the $\sigma_3$ colour structures can be safely neglected
even for the case of the model of ref. \cite{[1]}. 
Such assumptions are retrospectively justified
by the 
fact 
that the model of \cite{[1]} cannot yield supersymmetric points
even under the above assumption, 
for other reasons to be discussed below. 
Thus our approach in this 
paper is to identify the circumstances under which 
deformations of the 
model presented in \cite{[1]} can yield such points in the parameter space.

Notably, the situation $J >> t_+'$ may be met in the models 
of Dagotto {\it et al.}~\cite{[1]}, where NNN hopping $t' $ 
is neglected, but where the Coulomb attraction (\ref{2.5a}) 
is present, in order to guarantee the existence of 
$d$-wave superconducting gaps~\cite{dagottorev}.
Moreover, in the context of generalizations 
of the 
$t-V-J$ models of  
Feiner {\it et al.}~\cite{[1]},
such a situation (c.f. (\ref{param})) is 
met if one assumes an appropriate attractive $V'$,  
of opposite sign to the repulsion $U_{eff}$,
but close to it in magntitude
(notice that, on account of (\ref{2.6b}), in our generalization fo the t-V-j
model, one 
should replace $t$ in (\ref{param}) by $t_+'$). 
In such a case 
one has 
an additional large dimensionful scale $U_{eff}$, like in the case of 
the conventional Hubbard model of \cite{[5]}.

We next remark that 
in conventional non-abelian gauge theories the 
fermionic fields are usually spinors in the fundamental 
representation of the gauge group. 
Let us examine under what condition this is feasible in our case.
To this end we 
assemble the fermionic degrees of freedom into two 2-component
Dirac spinors~\cite{[5]}:
\be
{\tilde \Psi}_{1,i}^\dagger =\left(\psi_1~~-\psi_2^\dagger\right)_i,~~~~
{\tilde \Psi}
_{2,i}^\dagger=\left(\psi_2~~\psi_1^\dagger\right)_i\label{2.5}
\ee
where $\alpha$ in ${\tilde \Psi }^\dagger_{\alpha,i}$ is the colour index. 
We also consider very weakly coupled $SU(2)$ gauge groups, 
with couplings $g_{\rm SU(2)} \equiv g_2 << 1$.  
In the 
weak gauge field approximation, where 
the gauge group element (link) along the $\mu$ space-time direction is 
$U_{ij;\mu} \sim \int_i^{j;\mu} B_\mu^a \sigma^a + {\cal O}(g_2^2)$
(with $\sigma^a, ~a=1,2,3$ the Pauli matrices),   
one observes the following mathematical identities: 
\bea
&~& {\rm Tr}\left(\psi_i \psi^\dagger _{i+\mu}\right) = 
{\tilde \Psi}^\dagger _{i}{\tilde \Psi} _{i+\mu} \nn \\
&~&  {\rm Tr}\left(\psi_i \sigma_1 \psi^\dagger _{i+\mu}\right) = 
{\tilde \Psi}^\dagger _{i} \tau_1 {\tilde \Psi} _{i+\mu} \nn \\
&~& {\rm Tr}\left(\psi_i \sigma_3 \psi^\dagger _{i+\mu}\right) = 
{\tilde \Psi}^\dagger _{i} \tau_3 {\tilde \Psi} _{i+\mu} \nn \\
&~& {\rm Tr}\left(\psi_i \sigma_2 \psi^\dagger _{i+\mu}\right) = 
i\left(-{\tilde \Psi}^\dagger _{i} \sigma_3 \frac{1}{2}(\tau_1 + i
\tau_2){\tilde \Psi} _{i+\mu} 
+ {\tilde \Psi}^\dagger _{i}\frac{1}{2}(1 + \tau_3){\tilde \Psi} _{i+\mu}
\right)
\label{tausigmas}\eea
where the Pauli matrices $\tau^a,~a=1,2,3$ refer to `colour' space, 
and should be distinguished from the $\sigma_3$ matrices, which 
although are `colour' matrices, they refer to the action (\ref{1.6}),
in which the fermionic degrees of freedom consist of Grassmann variables
assembled in $2 \times 2$ matrices. 
From the last of (\ref{tausigmas}), therefore, it becomes evident
that the action (\ref{1.6}), may be mapped to a conventional 
lattice action, with spinors (\ref{2.5}) in the fundamental 
representation of the `colour' group, provided that the 
coupling $g_2 << 1$ is {\it weak}, 
and in addition there is a {\it gauge fixing}~\footnote{Note that 
the requirement for weak $g_2$ coupling is essential, given the fact that 
due to the non-Abelian nature of the gauge field, the local gauge 
fixing $B_\mu^2 =0$ alone is not sufficient to eliminate dangerous terms 
proportional to $\sigma^2$; this can be easily seen from the 
Bekker-Hausdorff identity:
$e^{ig_2 \sum_{a=1,3} \sigma^a B_\mu^a} = 
\left(\Pi_{a=1,3} e^{ig_2 \int _i^j \sigma^a B_\mu^a}\right) 
e^{\frac{1}{2}(g_2)^2 [\sigma^1,\sigma^3]B_\mu^1 B_\mu^3 + \dots }$,
with the commutator being proportional to $\sigma_2$; however 
such terms are of higher order in $g_2$, and hence restriction to 
weak couplings suffices to yield the conventional relativistic 
gauge form of the effectvie action
upon the appropriate gauge fixing.}: 
\be
\int_i^j dx^\mu B_\mu^2  =0~.
\label{gf}\ee
The weakness of the $SU(2)$ coupling guarantees that a 
mass gap in the problem is only generated by the $U_S(1)$ 
group~\cite{[5]}. In the context of the Hubbard model 
of \cite{[5]},
the coupling $g_2$ of the gauged $SU(2)$ 
interactions, pertaining to the spin degrees 
of freedom in the problem, is naturally weak, since it is 
related to 
the Heisenberg exchange energy $J$. Given that 
in three space-time dimensions the gauge couplings are dimensionful,
with dimensions of energy, one may define dimensionless couplings
by dividing them with the ultraviolet scale of the low-energy theory,
which in the model of \cite{[5]} is the (strong) Hubbard interaction $U >>
J$.
Thus a dimensionless coupling $g_2 \sim J/U << 1$ is naturally small 
in this context. A similar situation arises in the context of 
the effective single-band t-V-j model of \cite{[1]}, in the large 
$U_{eff} >> J$ limit (c.f. (\ref{param})).
On the other hand, the strong $U_S(1)$ coupling 
$g_1$, responsible for mass gap generation 
for the holons, may be assumed to be of order $U_{eff}$,
since this is the highest energy scale. However, in 
general for $t-j$ models that we consider these relations may not
be valid. Still as we shall see below, the 
ultraviolet cut-off of the effective theory, in the regime
relevant for supersymmetric points, we are interested here,  
may be up to two orders
of magnitude higher than $J$, thereby allowing the $U_S(1)$ 
interactions to be considerably stronger than the $SU(2)$ ones, 
if one wishes so.

To generate the conventional Dirac $\gamma$-matrix structure
for the fermionic action one may redefine the spinors
in the path integral ${\tilde \Psi} \rightarrow \Psi$, 
where   
$\Psi$ are  
{\it two-component} `coloured' 
spinors,
related to  the spinors in (\ref{2.5}) 
via a Kawamoto-Smit transformation~\cite{[20]}
\be
\Psi _\alpha(r) =\gamma_0^{r_0}\dots \gamma_2^{r_2}{\tilde \Psi} _\alpha(r)
\qquad  {\overline \Psi} _\alpha(r) 
={\overline {\tilde  \Psi }}_\alpha(r)(\gamma_2^{\dagger})^{ r_2}\dots 
(\gamma_0^{\dagger})^{r_0}\label{2.8}\ee
where $r$ is a point on the euclidean lattice,  
and $\alpha =1,2$ is a  `colour'
index, expressing the initial 
antiferromagnetic nature of the system.
The generation of Dirac $\gamma$ matrices
follows from identities of the form:
\be
   {\tilde \Psi}^\dagger _i \Delta {\tilde \Psi}_{i+\mu} 
= \Psi^\dagger _i \gamma_\mu \Delta \Psi _{i+\mu} (-1)^{\mu+\mu_2 (i_0 +
i_1) + \mu_1 i_0}~, \qquad \Delta = 1, \tau_1, \tau_3  
\label{ksident} 
\ee
but again the terms proportional to $\sigma_2$ ($\tau_2)$ are problematic
and can be eliminated by virtue of the gauge fixing (\ref{gf}). 
The 
$\gamma$ matrices appearing in (\ref{ksident}) 
are $2 \times 2$ antihermitean Dirac 
matrices on a Euclidean Lattice satisfying the 
algebra 
\be
\{ \gamma_\mu~,~\gamma _\nu \} =-2\delta _{\mu\nu}\label{2.9}\ee
In terms of the Pauli matrices $\sigma_i,i=1, \dots 3$, the 
$\gamma$ matrices are given 
by $\gamma _\mu = i\left( {\sigma _3,\sigma _1,\sigma _2} \right)
$. N.B. that 
fermion bilinears of the form  
${\overline \Psi}_{i,\alpha} \Psi _{i,\beta}$ ($i$=Lattice
index) 
satisfy 
\be
{\overline \Psi}_{i,\alpha} \Psi _{i,\beta} = 
{\overline {\tilde \Psi}}_{i,\alpha} {\tilde \Psi}_{i,\beta}
\label{psipsitilde}
\ee
due to the 
Clifford algebra (\ref{2.9}), 
and (anti-) hermiticity properties 
of the $2\times 2$ $\gamma$ matrices
on the Euclidean lattice. As we shall see later on,
this last identity will be crucial in yielding
a relativistic form of the 
effective action for the interacting nodal liquid of excitations 
in generalized Hubbard models. 

We next notice that 
on a lattice, in the path integral over
the fermionic degrees of freedom in a quantum theory, 
the variables ${\overline \Psi}$ and $\Psi$ are viewed as {\it independent}.
In view of this, 
the spinors $\Psi^\dagger_\alpha $ 
in (\ref{2.5})
may be replaced by ${\overline \Psi _\alpha}$,     
as being path integral variables
on a Euclidean Lattice appropriate for the Hamiltonian 
system (\ref{1.1}). 
This should be kept in mind when discussing
the microscopic structure of the theory in terms of the 
holon creation and annihilation operators $\psi_\alpha^\dagger, \psi_\alpha,
\alpha=1,2$.

An issue that should be dealt with properly is the appearance of 
the factors \\ 
$(-1)^{\mu + \mu_2 (i_0 + i_1) + \mu_1}$
in (\ref{ksident}), which would prevent the 
conventional Dirac structure to emerge.
However, this problem is easily arranged by absorbing such factors
in the quantum fluctuations of the electromagnetic field 
$U(1)^{em}$, which are integrated in a path 
integral~\footnote{A mathematically equivalent, but physically 
different way, of course, would be to assume 
a flux phase background for the electromagnetic field 
with flux $\pi$ per lattice 
plaquette~\cite{[2]}, and then consider quantum fluctuations 
about it in a path integral. This would wash out any remnant of the 
flux phase from the effective action, 
but help in absorbing the above-mentioned factors in a 
physically irrelevant 
normalization.}.
Notice that by doing so, one does not disturb the form of the 
bosonic $CP^1$ parts of the effective action, given that the 
$z$ magnons are electrically neutral. 
It is understood of course, that for the purposes of this work,
we shall not be interested further in the quantum fluctuations 
of the electromagnetic field, as their coupling is really much weaker than 
the couplings of the statistical gauge groups under consideration.
From now on, the electromagnetic interaction will be treated only as  
an external background.

The fermionic part of 
the long-wavelegth lattice lagrangian, then, reads:
\bea
&~& S={1\over 2}K' \sum_{i,\mu}[{\overline \Psi}_i (-\gamma_\mu) 
U_{i,\mu}V_{i,\mu} \Psi_{i+\mu}  + \nn \\
&~& {\overline \Psi}_{i+\mu}
(\gamma _\mu)U^\dagger_{i,\mu}V^\dagger_{i,\mu}
\Psi _i ]  + {\rm Bosonic~CP^1~parts}\label{2.6}
\eea
where the Bosonic $CP^1$ parts denote magnon-field $z$ dependent 
terms, and  
are given in (\ref{1.6}). It should be stressed once again
that this relativistic form is derived for a weakly-coupled
$SU(2)$ gauge group, and under a specific gauge fixing.
However, in view of the gauge invariance, characterizing 
(\ref{1.6}) and (\ref{2.6}), the physical results based 
on the above effective actions, in particular 
the existence of supersymmetric points in the 
parameter space, of interest to us here, 
are independent of the the gauge chosen.  

An additional point we would like to make concerns 
the relativistic form of the action (\ref{2.6}).
Although in (\ref{2.6}) 
we did not give explicitly the $CP^1$ parts,
however we have tacitly assumed 
the {\it equality of the effective velocities}
for spin $v_S$ and charge $v_F$ (Fermi velocity of holes) 
degrees of freedom. 
If such an assumption 
is not made, then the relativistic invariant form of  
of the effective lagrangian is spoiled~\cite{[2]}.
This can be easily understood by the 
fact that
in the effective lagrangian  
(obtained as a Legendre transform from the appropriate Hamiltonian)
the (different) velocities $v_S$ and $v_F$ 
enter in the derivatives with respect to the time variable,
e.g. $\partial/v_S \partial t$ ($\partial /v_F \partial t$) 
in the respective kinetic terms 
for spinons 
(holons). 
However, at the supersymmetric points of the 
nodal liquid, where, as we shall discuss 
later on, the dynamically-generated mass gaps between spinons and holons
must be equal, the equality $v_S=v_F$ is essential, otherwise
there would be different dispersion relations, leading to 
a difference in mass gaps. These comments should be understood
in what follows. From now on we shall work in units 
of the fermi velocity $v_F$. 

The coefficient $K'$ is a constant 
which stems from the $t_{ij}-$ and $\Delta _{ij}-$ dependent coefficients
in front of the fermion terms in (\ref{1.6}). 
An order of magnitude estimate of the  
modulus of (the shifted) 
$\Delta '_{ij}$ then, which determines the strength of the 
coefficient $K'$, may be provided by its equations of motion. 
Assuming that the modulus of (the dimensionless) 
fermionic bilinears is of order unity, then, we have as an order of
magnitude 
\be
K' \sim \left(t'_{+} + {J \over 8}\right)
\label{anonymous}
\ee 
Note that in the regime of the parameters of \cite{[1]} 
$t << t_+$ and $t_+ \simeq {3 \over 2}J$ 
for momenta close to a node in the Fermi surface, 
of interest to us here. Thus 
\be
K' \simeq 25J/8 \label{2.6c}\ee 
However, 
one may even 
consider more general models, in which $K'$ and 
the Coulomb intersite interaction 
$V$ are treated as independent phenomenological parameters.

\subsection{Field-Theoretic Treatment of the Constraints}

As discussed in Appendix B, supersymmetrization 
of $CP^1$ type models, like the ones considered here, 
requires that the $CP^1$ constraint be of the form 
$\sum_{\alpha=1}^{2}|z_\alpha |^2 =1$. 
In our case, however, 
the no-double 
occupancy 
constraint, when expressed in terms of the $z$ 
and ${\tilde \Psi} _\alpha$, $\alpha=1,2$, (spinor) fields,
with $\alpha$ a `colour' index, is written as: 
\be
\sum_{\alpha=1}^{2} [ {\overline z}^\alpha z _\alpha + \beta   
{\overline {\tilde \Psi}}^{\alpha} \sigma _3 {\tilde \Psi} _\alpha ] = 1 
\label{B.2b} 
\ee
where $\beta = 1/K'^2$, $K'$ is given by (\ref{2.6c}), 
the $2 \times 2$ matrix 
$\sigma _3 $ acts in spinor space, and  
the fermions ${\tilde \Psi}$ are the  
{\it two-component} spinors (\ref{2.5}). 
Equivalently the fermion bilinear terms 
in (\ref{B.2b}) can be expressed in terms of the spinors $\Psi$ (\ref{2.8}),
which have conventional  
Dirac kinetic terms. This is due to identities 
of the form (\ref{ksident}), 
extended appropriately to the case of coincidence limits
in the continuum formalism. The relevant  
$(-1)$ factors in that case 
may be absorbed in the definition of $\beta$. 
It is understood that 
appropriate rescalings can be made 
in the definition of the spinors so as to ensure the canonical 
kinetic (Dirac) term. We have also taken into account 
that in a euclidean path-integral 
the variables $\Psi^\dagger$ and $\Psi$ are viewed 
as independent, which implies that one may redefine 
$\Psi^\dagger \rightarrow {\overline \Psi}$ where ${\overline \Psi}$
in later analysis will nevertheless be considered in the conventional way,
i.e as ${\Psi ^\dagger}\gamma _0$. Consequently we can interpret the fermion
term in the constraint (\ref{B.2b}) as $\Psi^\dagger \Psi$ the fermion
number term.

The presence of the $\Psi ^\dagger \Psi $ (non-relativistic) 
fermion number term in the constraint (\ref{B.2b})  
appears at first sight to complicate things, since  
the conventional $CP^{1}$ constraint 
$|z|^2 = 1$ is no longer valid.
In fact, as discussed in Appendix B, supersymmetry 
is compatible with the following form of the constraints:
\be
|z_\alpha |^2 = 1~, \qquad {\overline z}_\alpha \Psi _\alpha = 0
\label{cp1constraint}
\ee
arising from the superfield version of the $CP^1$ 
constraint~\cite{[8],[11]}. In fact the fermionic 
counterpart of (\ref{cp1constraint}) 
can be solved by means of a `colourless' 
fermion field ${\cal X}$ 
that satisfies (on account of the bosonic $CP^1$ parts 
of (\ref{cp1constraint})):
\be
   \Psi_\alpha = \epsilon_{\alpha\beta}{\overline z}_\beta {\cal X}~,
    {\cal X}=\epsilon_{\alpha\beta}z_\alpha\Psi_\beta
\label{solution}
\ee
where $\Psi_\alpha$ are the Dirac spinors defined above. 
To ensure the conventional $CP^1$ form of the bosonic 
part of the supersymmetric constraints 
(\ref{cp1constraint}) from (\ref{B.2b}) 
we should demand $\beta << 1$, which is satisfied 
in a regime of the parameters of the theory for which 
\be
K' >> K = \sqrt{J}\left|\Delta _z \right|\left(1 - \delta \right),\qquad 
0 < \delta < 1   
\label{2.6d}\ee 
For the model of \cite{[1]}, for instance, 
on account of (\ref{2.6c}), this condition implies
that 
\be
\sqrt{J}/\left|\Delta _z \right| \gg 0.32~(1-\delta)~,\qquad 
0 < \delta < 1   
\label{cond2}
\ee
By 
appropriately rescaling the fermion fields $\Psi$ to $\Psi'$, so that in the
continuum they have a canonical Dirac term, 
we may 
effectively constrain
the $z$ fields to satisfy the 
$CP^1$ constraint:
$$ |z_\alpha |^2 + {1\over K'} ({\rm \Psi'-bilinear~terms}) 
=  1$$
where now the fields $\Psi$ are dimensionful. 
with dimensions of 
$[energy]$. A natural order of magnitude of these dimensionful 
fermion bilinear 
terms is of the order of $K^2$, which plays the r\^ole of the characteristic
scale in the theory, being related directly to the 
Heisenberg exchange energy $J$.  
In the limit $K' >> K$ (\ref{2.6d}) therefore  the fermionic terms in the 
constraint can be  ignored, and the constraint assumes the standard 
$CP^1$ form involving only the $z$ fields ( this being also the case 
for the model of \cite{[2],[3]},
in a specific regime of the 
microscopic parameters).  

As we shall see later, however, 
the condition (\ref{cond2}) alone, although 
necessary, is not sufficient to guarantee 
the existence of supersymmetric points.  
Supersymmetry imposes 
additional restrictions,  
which in fact rule out 
the existence of supersymmetric points for the 
model of \cite{[1]} compatible with 
superconductivity~\footnote{We note in 
passing that in realistic materials superconductivity 
occurs for doping concentrations above $3\%$, and is destroyed 
for doping concentrations larger than $\delta _{\rm {max}} \sim 10\%$.}. 
However, this does not prevent 
one from considering more general models 
in which 
$K'$ is viewed as a phenomenological parameter, not constrained 
by (\ref{2.6c}). In that case, supersymmetric points may occur 
for a certain regime of the respective parameters.

However, as a result of the spin-charge separation 
formalism, there is a different way to treat the constraints
in a pth integral, which however takes into account 
the coupling of the system to an external elelctromagnetic field, and as
such
is not apriori relevant to the supersymmetric regime.
Nevertheless, as we shall discuss in section 6, this will be relevant for 
electric charge transport in the model for which supersymmetry
(in the absence of external fields) will be argued to 
play a rather crucial but sabtle r\^ole. 

Indeed we observe that 
the fermion number terms in (\ref{B.2b})
may be absorbed in a rescaling of the (quantum fluctuations of the) 
temporal component of the electromagnetic field $A_0({\vec x}, t)$,
which couples (relativistically) only to the spinors  $\Psi$ 
(see section 6.2 below). Indeed, by implementing
the constraint (\ref{B.2b}) in a path-integral 
via the introduction of a Lagrange multiplier field $\lambda (x)$:
\be
    \delta (|z_\alpha |^2 + \beta {\overline \Psi}_\alpha \sigma_3 \Psi -1)=
\int D\lambda (x) e^{i\lambda (x) \left(z_\alpha {\overline z}_\alpha +  
\beta {\overline \Psi}_\alpha \sigma_3 \Psi -1\right)}
\label{lagrangemult}
\ee
Upon absorbing $\lambda(x)$ in a shift of $A_0 ({\vec x},t)$,
one obtains from the Maxwell terms in the electromagnetic part 
of the effective action the following combination:
\be
   {\cal L}_{em} \ni -\frac{1}{4(e^2/c^2)}
\left(2\partial_i \lambda F_{0i} + (\partial_i \lambda )^2
\right) + {\rm standard~~Maxwell~~terms}
\label{maxwellact}
\ee
where $F_{0i}$ is the appropriate components 
of the Maxwell tensor of the 
(redefined) electromagnetic field, the index $i$ 
is a spatial index, and repeated indices denote summation, 
The equations of motion for $\lambda $ in the effective action 
obtained
after integrating out, say, the $z$ degrees of fredom 
yield the standard $CP^1$ model terms~\cite{polyakov}, but also
terms of the form $\nabla_i^2 \lambda + 2\nabla^i F_{0i} $.
One, therefore, may consider 
a phase in which $<\lambda (x) > = {\rm const} \ne 0$,
provided that the electromagnetic field is chosen as an external one, 
satisfying Maxwell's equations, which is our case.

The Bosonic part of the constraint, then, implies a mass
for the spinons $m_z \propto <\lambda (x)>$~\cite{polyakov}.
The fermionic part on the other hand has the form of a 
temporal component of the electric current (see section 6.2 below). 
The coefficient $\beta <\lambda (x)>$ may be absorbed in a shift
of the quantum fluctuations of $A_0 ({\vec x},t)$. As already stated
previously,
quantum fluctuations of the electromagnetic field will not be of 
further interest
to us here, given that we shall treat it only as external background. 

From the above discussion it becomes clear, then, that in either case 
one maps the double occupancy constraint 
(\ref{B.2b}) into the standard $CP^1$ constraint:
\be 
    \sum_{\alpha=1}^{2} \left| z_\alpha \right|^2 = 1 
\label{3.7b}
\ee
However, as we have explained above, one cannot really avoid
the restriction (\ref{cond2}), as far as the existence of supersymmetric
points is concerned, 
given that any alternative treatment would require coupling the system 
to (supersymmetry-breaking) 
external electromagnetic fields, 
since otherwise the fermionic parts of (\ref{B.2b}) would be present. 
As we shall see in section 6, though, the alternative treatment
of the constraint leads to interesting phases of the theory
characterized by superconducting elelctric-charge transport. 
And, then, any supersymmetry that might have existed before 
coupling to elelctromagnetism would play an important (but subtle)
r\^ole in ensuring the existence of 
superconductivity. 
 
In addition to the $CP^1$  constraint, 
one also encounters the remaining of the constraints (\ref{3.3}),
which may also be treated using appropriate Lagrange multiplier
fields $\lambda_2(x), \lambda _3(x)$ representations 
for the respective $\delta$-functionals 
$\delta(\psi^\dagger_{1,i}\psi^\dagger_{2,i})$,
$\delta(\psi_{1,i}\psi_{2,i})$:
\bea
&~&\delta (\psi_1 \psi_2)\delta(\psi^\dagger_1 \psi^\dagger_2)  \sim  
\int d\lambda_2 d\lambda_3(x) 
 e^{i\int d^3x \lambda_2(x) \psi_1 (x) \psi_2 (x) + 
 i\int d^3x \lambda_3(x) \psi_1^\dagger (x) \psi_2 (x)^\dagger} 
\propto \nn \\
&~&\int d\lambda_2(x)d\lambda_3(x) 
e^{i\int d^3x \left[\frac{\lambda_2 (x)-\lambda_3(x)}{2} 
\sum_{\alpha} {\overline \Psi}^\alpha \gamma_1 \Psi_\alpha 
+\frac{\lambda_3 (x)}{2}\sum_{\alpha} {\overline \Psi}^\alpha 
\gamma_2 \Psi_\alpha \right]}
\label{deltident2}
\eea 
Above we have expressed the relevant constraint in terms
of the spinors (\ref{2.8}), using the identity (\ref{ksident}),
appropriately applied to the case of 
coincident limits in the continuum formalism, 
and absorbed relevant $(-1)$ factors in redefinitions
of the lagrange multiplier fields $\lambda_i,~i=2,3$. 
Notice that the spatial $\gamma_j$,
$j=1,2$ Dirac matrices are 
expressed in terms of the $2\times 2$ off-diagonal 
Pauli matrices $\sigma_j$,$j=1m2$ as $\gamma_j=i\sigma_j$.  
To obtain information about the new phases it is necessary to assume 
$<\lambda_2(x)>,<\lambda_3(x)> \ne 0$.  
We thus observe that the structures in (\ref{deltident2}) 
resemble terms pertaining to 
``electric current'' operators 
 $J_i = {\overline \Psi}\gamma_i \Psi$, $i=1,2$ (see sec. 6.2), and as such 
can be absorbed in the quantum fluctuations of the spatial components
of the electromagnetic field ${\vec A}(\vec x,t)$. 

It should be stressed again that the situation in which 
the Lagrange multiplier fields acquire non-zero vacuum expectation
values (vev),
$<~\lambda(x)~>,<~\lambda_i>\ne0,~i=2,3$, corresponds to the selection 
of a specific 
ground-state of the system (phase), about which one considers 
quantum fluctuations. There is always the phase 
in which such vev's are zero, in which case one implements the constraints
directly on the path-integral correlators, e.g. correlation fucntions 
proportional to $\psi_1\psi_2$  are set to zero in this phase.
In what follows, first we shall 
resolve the constraints
in this latter phase, and later on (section 6) we shall discuss
the other phases of the model. As we shall later, this phase
is characterized by spin transport but not electric charge transport,
a situation that should be compared with the case of the 
nodal liquids
of ref. \cite{fisher}
in the electrically-neutral-fermion representation for spinons. 
On the contrary, as we shall show in section 6, 
the phase in which the lagrange multiplier vev's are non trivial
may yield unconventional superconductivity of Kosterlitz-Thouless 
type~\cite{[2],[5]}.

With the above in mind we consider from now on the standard 
$CP^1$ constraint involving only $z$ fields. 
By an appropriate normalization of $z$ to $z' = {z \over \sqrt{1 - \delta}}$
the constraint then acquires the familiar normalized $CP^1$ form 
$|z_\alpha|^2=1$ form.
This implies a rescaling of the normalization coefficient $K$  
in (\ref{1.6}):  
\be
K \rightarrow  {1 \over \gamma} \equiv K (1-\delta) \simeq \sqrt{J}|\Delta
_z| (1-\delta)^2  
\label{3.6e}
\ee
In the naive continuum limit, then, the effective lagrangian 
of spin and charge degrees of freedom describing the low-energy 
dynamics of the Hubbard (or $t-j$) model (\ref{1.6}) of \cite{[5]} 
is then: 
\be
{\cal L}_2 \equiv  {1 \over \gamma }{\rm Tr}\left| \left( {\partial _\mu +
ig_2 \tau^a B_\mu^a + ig_1a_\mu } \right)z \right|^2 +
{\overline \Psi}D_\mu\gamma_\mu\Psi \label{3.7}\ee
with $z_\alpha$ a complex doublet satisfying the constraint 
(\ref{3.7b}). 
The Trace ${\rm Tr}$ is over group indices, 
$D_\mu = \partial_\mu -ig_1a_\mu^S-ig_2\tau^aB_{a,\mu}
-{e \over c}A_\mu$,
$B_\mu^a$ is the gauge potential of the local (`spin') $SU(2)$ group,
and $a_\mu$ is the potential of the $U_S(1)$ group.

It should be remarked that, 
we are 
working in units of the Fermi velocity $v_F (=v_D)$ of holes, which plays
the 
r\^ole of the limiting velocity for the nodal liquid. 
We stress
once again that for the nodal liquid at the supersymmetric points 
we have assumed that $v_F \simeq v_S$, where $v_S$ is the 
effective velocity of the spin degrees of freedom. 
The relativistic form of the fermionic and bosonic
terms of the action (\ref{3.7}) 
is valid {\it only} in this regime
of velocities. This is sufficient for our purposes in this work. 
Indeed, at the supersymmetric points, 
where we shall restrict our analysis here, 
the mass gaps 
for spinons and holons, which may be generated {\it dynamically},  
are {\it equal} by virtue of supersymmetry at zero temepratures and
in the absence of any external fields. Hence it makes sense
to assume the equality in the propagation velocities
for spin and charge degrees of freedom,
given that this situation is consistent with the 
respective dispersion relations. This is {\it not true},
of course, for excitations away from such points.

\section{The NN interaction terms $H_V$ }

We will now discuss the Coulomb-interaction (attractive) terms 
\be
H_V=-V_{total} \sum\limits_{\left\langle {ij} \right\rangle } {n_in_j}
\label{anon2}
\ee
introduced in ref. \cite{[1]}, where $V_{total}$ is given in (\ref{totalv}).
With the above discussion in mind 
for the spinors (\ref{2.5})
we note that, under the ansatz (\ref{3.1}), at a site $i$
the electron number operator  $n_i$ 
is expressed, through the Determinant (Det) of the $\chi$ matrix in 
(\ref{3.1}),  
in terms of the spin, $z_\alpha,\alpha=1,2$, 
and charge $\psi_\alpha, \alpha=1,2$, operators 
as: 
\bea
&~&n_i \equiv \sum_{\alpha = 1}^{2}
 c^{\dagger}_{\alpha,i}c_{\alpha,i}={\rm Det}\chi_{\alpha\beta,i} = \nn \\
&~&{\rm Det}{\hat z}_{\alpha\beta,i} + {\rm Det}{\hat \psi}_{\alpha\beta,i}
=
\sum_{\alpha=1}^2 \left( \psi_\alpha 
\psi_\alpha ^\dagger + |z_\alpha|^2 \right)
\label{4.1}\eea 
We may express the quantum fluctuations for 
the Grassmann fields  
$\psi_\alpha$ (which now carry a `colour' index $\alpha=1,2$
in contrast to Abelian spin-charge separation models) 
via: 
\be
\psi _{\alpha, i}\psi _{\alpha,i}^+=\left\langle {\psi _{\alpha, i}
\psi _{\alpha, i}^+} \right\rangle +  
:\psi _{\alpha, i}\psi _{\alpha, i}^+:\,,~{\rm no~sum~over~i}
\label{4.2}\ee
where $: \dots :$ denotes normal ordering of quantum operators,
and from now on, unless explicitly stated, 
repeated indices are summed over. 
Since $$\left\langle {\psi _{\alpha, i}\psi _{\alpha, i}^+} 
\right\rangle \equiv 1-\delta~,~~~{\rm no~sum~over~i}$$ 
$\delta$ the doping concentration 
in the sample (\ref{2.7}),  
we may rewrite $n_i$ as $$n_i = \left(|z_\alpha |^2 + (1 -\delta) 
+ :\psi_\alpha \psi_\alpha^\dagger: \right)_i$$
which in terms of the spinors ${\tilde \Psi}$ 
is given by (c.f. 
(\ref{2.5})): 
\be
n_i = 2 -\delta + {1 \over 2} 
\left({\tilde \Psi}^\dagger_\alpha \sigma_3 
{\tilde \Psi}_\alpha \right)_i\label{3.3b}\ee
where $\sigma_3 = \left(\begin{array}{cc}
1 \qquad 0 \nn \\
0 \qquad -1 \end{array}\right)$ acts in (space-time) 
spinor space,
and we took  
into account the $CP^1$ constraint (\ref{3.7b}).  

Consider now the attractive interaction term $H_V$ (\ref{anon2}), 
introduced 
in ref. \cite{[1]}. 
We then observe than 
the terms linear in $(2-\delta)$ in the expression for $H_V$  
can be absorbed by an appropriate shift 
in the chemical potential, about which we linearize to obtain the 
low-energy theory. We can therefore ignore such terms from now on.  
 
Next, we make use of the fact, mentioned earlier, 
that in a lattice path integral 
the spinors ${\tilde \Psi}^\dagger_\alpha $ 
may be replaced by ${\overline {\tilde \Psi} _\alpha}$.    
From the structure of the spinors (\ref{2.5}), then,  
we observe that we may rewrite the $H_V$ term 
{\it effectively} as
a Thirring vector-vector interaction among the spinors ${\tilde \Psi}$ 
\be
H_V=+{V_{total} \over 4} \sum_{<ij>} \left( {\overline {\tilde \Psi}}_\alpha
\gamma _\mu {\tilde \Psi} _\alpha \right)_i
\left( {\overline {\tilde \Psi}}_\beta \gamma ^\mu {\tilde \Psi} _\beta
\right)_j \label{4.5}\ee
where summation over the repeated indices $\alpha, \beta (=1,2)$, 
and $\mu=0,1,2$, with $\mu=0$ 
a temporal index,
is understood. To arrive 
at (\ref{4.5}) we have expressed $\sigma_3$ as $-i\gamma_0$, and  
used the Clifford algerba (\ref{2.9}),
the off-diagonal nature of the $\gamma_{1,2}=i\sigma_{1,2}$ matrices,
as well as the constraints
(\ref{3.3}). In particular the latter imply that any scalar product 
between Grassmann variables $\psi_\alpha$ (or $\psi^\dagger_\beta$)  
with different `colour' indices {\it vanish}. 

Taking the  continuum limit of (\ref{4.5}), and ignoring 
higher derivative terms involving four-fermion 
interactions, which by power counting are irrelevant 
operators in the infrared, 
we obtain after passing to a Lagrangian formalism
\be
{\cal L}_V = 
-{V_{total} \over 4 K'^2} \left( {\overline {\tilde \Psi} _\alpha} \gamma
_\mu {\tilde \Psi} _\alpha
\right)^2 \label{4.6}\ee
where we have used rescaled spinors, with 
the canonical Dirac kinetic term with unit coefficient,  
for which the canonical form of the $CP^1$ 
constraint (\ref{3.7b}) 
is satisfied. For notational convenience 
we use the same notation ${\tilde \Psi}$ for these spinors as the unscaled
ones. Although this 
is called the naive continuum limit, it actually captures correctly the
leading infrared behaviour of the model.

We then use a Fierz rearrangement formula for the $\gamma$ matrices  
$$\gamma ^\mu _{ab}\gamma _{\mu,cd} = 
2 \delta _{ad}\delta _{bc} - \delta _{ab}\delta _{cd}$$
where Latin letters indicate spinor indices, and Greek Letters
space time indices.  
The Thirring (four-fermion) 
interactions (\ref{4.5}) then 
become: 
\be
\left({\overline {\tilde \Psi}}_\alpha \gamma _{\mu} {\tilde \Psi} _\alpha
\right)^2= 
-3 \left({\overline {\tilde \Psi}} _\alpha {\tilde \Psi} _\alpha \right)^2 
- 4 \sum_{\alpha < \beta} 
\left({\overline {\tilde \Psi} }_\alpha {\tilde \Psi} _\beta  {\overline 
{\tilde \Psi} }_\beta 
{\tilde \Psi} _\alpha \right)  
\label{4.7}\ee
Notice that this form permits us to use, on account of the 
identity (\ref{psipsitilde}),  
either of the forms (\ref{2.8}) or (\ref{2.5}) 
for the spinors $\Psi$ or ${\tilde \Psi}$ in the expression 
of $H_V$.
It should be noted, though,  that the canonical Dirac form of 
the kinetic terms for the spinors is valid only in the form 
(\ref{2.8}), which we stick to from now on. 

As mentioned above, 
in the model of \cite{[5]}, due to the first of the constraints
(\ref{3.3}), the mixed colour terms vanish, thereby leaving us 
with pure Gross-Neveu {\it attractive} interaction terms of the form: 
\be
{\cal L}_V = +{3 V_{total} \over 4 K'^2} \left({\overline \Psi} _\alpha \Psi
_\alpha \right)^2
\label{3.8}\ee
which describe the low-energy 
dynamics of the interaction (\ref{anon2}) in the context 
of the non-Abelian spin-charge separation (\ref{3.1}). 
It should be
stressed that (\ref{3.8}) is specific to our spin-charge 
separation model.

Moreover in the context of the spinors (\ref{2.5}), a 
condensate of the form $<{\overline \Psi}_\alpha \Psi_\alpha>$ 
on the lattice {\it vanishes}
because of the constraints (\ref{3.3}). 
Such condensates would violate parity (reflection) 
operation on the planar spatial lattice, which on the spinors ${\tilde
\Psi}$ 
is defined to act as follows: 
$${\tilde \Psi}_1 \left( x \right) \rightarrow \sigma_1 {\tilde
\Psi}_2\left( x \right),\quad 
{\tilde \Psi}_2\left( x \right) \rightarrow \sigma_1 {\tilde \Psi}_1\left( x
\right)$$
or equivalently, 
in terms of the (microscopic) holon operxtors $\psi_\alpha, \alpha=1,2,$:
$$\psi_1\left( x \right) \rightarrow \psi_2^\dagger\left( x \right), \qquad 
\psi_2\left( x \right) \rightarrow -\psi_1^\dagger\left( x \right).$$   

To capture correctly this fact in the context of our effective continuum 
Gross-Neveu interaction (\ref{3.8}) 
the coupling strength {\it must} be subcritical,
i.e. weaker than the critical coupling for mass generation.  
As discussed in 
Appendix A, the critical coupling of the Gross-Neveu interaction is 
expressed in terms of a high-energy cut-off scale $\Lambda$ as~\cite{[9]}: 
\be
1=4 g_c^2 \int\limits_{S_\Lambda } {{{d^3q} \over {8\pi^3 q^2}}} = {2 g_c^2
\Lambda \over \pi ^2}
\label{3.9}
\ee
where $q$ is a momentum 
variable and ${S_\Lambda }$ is a sphere of radius $\Lambda$. The divergent
$q$-integral is cut-off at a momentum scale $\Lambda$ 
which defines the low-energy theory of interest. 
For the case of interest $g^2={3 V_{total} \over 4 K'^2}$; on 
using (\ref{2.6c}), then, the condition of sub-criticality requires that
\be
\Lambda  \lsim 77~J~.
\label{3.10}
\ee 
which is in agreement with the fact that in all  
effective models for doped antiferromagnets used in the 
literature
the Heisenberg exchange energy $J \sim 1000~{\rm K}$ serves as
an upper bound for the energies of the 
excitations
of the effective (continuum) theory. 
However, as mentioned above, 
to obtain a relativistic gauge theory from the 
lattice action (\ref{1.6}) one needs the $SU(2)$ interactions
to be considerably weaker than the $U_S(1)$ interactions, responsible 
for mass generation:
the above condition (\ref{3.10}) is also compatible with this, provided
one identifies the (dimensionful) coupling 
of the $U_S(1)$ interactions with a (high-energy) 
cut-off scale $\Lambda \sim 77~J$. In the context of the 
effective single-band t-V-j models (\ref{param}), for instance, 
$\Lambda$ may be identified with a $U_{eff} >> J$.  

\section{Dynamical Spinon-Holon Symmetry (Supersymmetry) 
in the Nodal Liquid and Potential 
Phenomenological Implications} 

\subsection{Conditions for N=1 Supersymmetry in the nodal liquid}

We turn now to conditions for supersymmetrization of the 
above continuum theory, i.e. conditions for dynamical symmetries
between the spinon (boson) and holon (fermion) degrees of freedom. 
Below we shall only outline the main results. 
Some technical details on the formalism are given in \cite{[8]}
and reviewed in Appendix B.
Since it has been argued that $U_S(1)$ is responsible for dynamical mass
generation (and superconductivity) in the model of \cite{[5]} we
shall ignore the non-Abelian $SU(2)$ interactions, keeping only the 
Abelian However since the latter argument is not rigorous, it would be
desirable to supersymmetrise the full group in order to check the phenomenon
of dynamical mass generation. The extension to supersymmetrizing the full
gauge multiplet 
$SU(2) \times U_S(1)$ will be the topic of 
a forthcoming work. However we shall still maintain 
the colour structure in the spinors, which is important 
for the ansatz (\ref{3.1})~\footnote{Ignoring the 
$SU(2)$ interactions implies, of course, that the `colour' structure
becomes a `flavour' index; however, this is essential  
for keeping track of the correct degrees of freedom 
required by supersymmetry in the problem at hand~\cite{[8]}.}. 

As discussed in detail in \cite{[8],[11]}, and reviewed briefly 
in Appendix B,  
the conditions for $N=1$ 
supersymmetric extensions of a $CP^1$ $\sigma$ model
is that 
the constraint is of the standard $CP^1$ form (\ref{3.7b}), 
supplemented by {\it attractive} 
four-fermion interactions of the Gross-Neveu type 
(\ref{3.8}), whose coupling is related to the 
coupling constant of the kinetic $z$-magnon terms of the 
$\sigma$-model in a way such as to guarantee the balance 
between bosonic and fermionic degrees of freedom 
Specifically, 
in terms of component fields, the pertinent lagrangian reads:
\be
L = g_1^2 [D_{\mu} \bar{z}^{\alpha}D^{\mu}z^{\alpha} +
i {\overline \Psi} \not{D} \Psi + \bar{F}^{\alpha} F^{\alpha} 
 + 2i({\overline \eta} \Psi^{\alpha} \bar{z}^{\alpha} -
{\overline \Psi}^{\alpha} \eta z^{\alpha})]
\label{B.16a} 
\ee
where $D_\mu$ denotes the gauge covariant derivative with respect to the 
$U_S(1)$ field. 
The analysis of \cite{[8],[11]} reviewed in Appendix B shows that, upon
using the equations of motion, 
\be
{\overline F}^\alpha F_\alpha = \sum_{\alpha=1}^{2} {1 \over 4} \left( 
{\overline \Psi}^\alpha \Psi_{\alpha} \right)^2 
\label{B.19a}
\ee
We thus observe that the $N=1$ supersymmetric extension 
of the $CP^1$ $\sigma$ model {\it necessitates} 
the presence of {\it attractive} Gross-Neveu type interactions among the 
Dirac fermions of {\it each sublattice}, 
in addition to the gauge interactions. 

In the context of the effective theory (\ref{3.7}), (\ref{4.6}), 
discussed 
in this article,
the $N=1$ supersymmetric
effective lagrangian (\ref{B.16a}) is obtained 
under the following restrictions among the 
coupling constants of the statistical model: 
\be
g_1^2={3 V_{total} \over K'^2} = \gamma = {1 \over \sqrt{J} |\Delta _z| 
(1 -\delta )^2}~,\qquad 0 < \delta  < 1    \label{4.1b}\ee
Note that in the context of the model of 
ref. \cite{[1]},  
for which (\ref{2.6a}),(\ref{2.6c}) are valid,
the relation 
(\ref{4.1b}) 
gives the supersymmetric point 
in the parameter space of the model 
at the particular
doping concentration $\delta = \delta _s$:
\be
\left(1 - \delta _s\right )^2 \simeq {3.89 \sqrt{J} \over |\Delta _z|}~,
\qquad 0 < \delta _s < 1   
\label{4.2b}
\ee
According to the discussion in section 4, 
unbroken supersymmetry (which is valid only in the absence of external
elelctromagnetic fields) imposes an additional restriction
(\ref{cond2}).  
Then we observe that 
compatibility of (\ref{4.2b}) with (\ref{2.6d}),(\ref{cond2})
requires : 
$1-\delta_s \gg 1.25$, which 
implies that the model of \cite{[1]}
does not have supersymmetric points.

However, 
one may consider 
more general models 
in which $V$ and $K' \sim
t'_+ + J/8 $ 
are treated as independent phenomenological parameters
(c.f. (\ref{2.6ab})); in such a case
one can obtain regions of parameters
that characterize the supersymmetric points (\ref{4.1b}),(\ref{4.2b}) 
compatible 
with superconductivity. 

Some comments are now in order:
First, it is quite important to remark that 
in the model of \cite{[5]}, where the 
antiferromagnetic structure of the theory is encoded in a colour 
(non-Abelian) degree of freedom of the spin-charge separated 
composite electron operator (\ref{1.5}) on a single lattice geometry,  
there is a matching between the 
 bosonic ($z$ spinon fields) and fermionic ($\Psi$ holon fields)  
physical degrees of freedom, as required by supersymmetry,         
without the need for 
duplicating them by introducing ``unphysical'' degrees 
of freedom~\cite{[8]}. 

The gauge multiplet of the $CP^1$ $\sigma$-model also needs a supersymmetric
partner which is a Majorana fermion called the gaugino.  
As shown in \cite{[8]}, and reviewed in Appendix B,
such terms lead to an 
effective electric-charge violating interactions on the spatial planes,
given that the Majorana gaugino is a real field, and as such cannot
carry electric charge (which couples as a phase to a Dirac field). These
terms can be interpreted as the removal or addition of electrons due to
interlayer hopping.

Indeed, the gaugino $\eta$ terms in 
the supersymmetric lagrangian (\ref{B.16a})  
have the form: 
\be
   \int d\eta  e^{2g_1^2i \int d^3 x {\overline \eta} \Psi ^\alpha 
{\overline z}_\alpha
+ H.C.}
\label{absencepth}
\ee
and hence may 
be viewed heuristically as constituting a Majorana-spinor 
representation of the {\it absence}
of spin {\it and} charge at a site of the planar lattice 
system
To understand this, 
the reader is advised to make 
a comparison with the Grassmann $\chi, \chi ^\dagger $,
representation of a Wilson line (`missing spin' S ) in the 
treatment of {\it static holes} in refs. \cite{Sha,[2]}:
\be
   \int d\chi ^\dagger d\chi  e^{-iS \int dt \sum _{i} (-1)^i \chi _i
^\dagger 
\chi _i a_0 (i, t) } 
\label{spin}
\ee
where $a_0$ is the temporal component 
of the gauge potential of the $CP^1$ $\sigma$-model, describing 
spin excitations in the antiferromagnet. 
From this point of view, 
the existence of $N=1$ supersymmetry in the doped 
antiferromagnets necessitates {\it interplanar couplings}, through 
hopping of spin and charge degrees of freedom (electrons) 
across the  planes. In view of (\ref{4.1b})
such interlayer hopping is 
suppressed by terms of order $\sqrt{J}$.

Another important point we wish to make concerns the 
four-fermion attractive Gross-Neveu interactions in 
(\ref{B.16a}),(\ref{B.19a}). 
As discussed in detail in \cite{fmm},
if the coupling of such terms is supercritical, then a parity-violating 
fermion (holon) mass would be generated in the model.  
However, the condition (\ref{3.10}), which is valid in the 
statistical model of interest to us here, implies that the 
respective coupling is always subcritical, and thus there 
is no parity-violating dynamical mass gap for the holons, induced 
by the contact  Gross-Neveu interactions.  This leaves one with the 
possibility of {\it parity conserving} dynamical mass generation,
due to the statistical gauge interactions in the model~\cite{[5],fmm}. 

A detailed analysis of such phenomena in the context
of our $CP^1$ model is left for future 
work. For the present, however, we note that 
in $N=1$ supersymmetric gauge models, supersymmetry-preserving 
dynamical mass is possible~\cite{[8],cm,cmp}. In fact,
as discussed in \cite{cmp}, although by supersymmetry 
the potential is zero, and thus there would naively seem that there is 
no obvious 
way of selecting the non-zero mass ground state
over the zero mass one, however there appear to be instabilities
in the {\it quantum effective action} in the massless phase, 
which manifest themsleves through 
instabilities of the pertinent running coupling.  
The opening of such a fermion mass gap 
has been associated with 
the existence of a non-trivial infrared fixed point 
of the renormalization-group flow, which implies 
non-fermi liquid behaviour~\cite{[21]}. 

From a physical point of view, such a phenomenon would imply 
that, for sufficiently strong gauge couplings, 
the zero temperature liquid of excitations 
at the nodes of a $d$ wave superconducting gap would be characterized  
by the dynamical opening of mass gaps for the holons. 
At zero temperature, and for the specific doping 
concentrations corresponding to the supersymmetric 
points, as advocated above, the nodal gaps between spinon and holons 
would be equal, in agreement with the assumed equality 
of the respective propagation velocities $v_F=v_S$,
which yielded the relativistic form of the effective continuum 
action (\ref{3.7}) of the nodal excitations
at the supersymmetric points. 

The opening of a nodal mass gap, due to the $U_S(1)$ gauge interactions,
would imply a breaking of the fermion number (global $U(1)$) symmetry,
and thus superconductivity upon coupling the system to external 
electromagnetic fields, 
according to the scenario of \cite{[2],[5]}, which is reviewed briefly
below for the benefit of the non-expert reader.

\subsection{Kosterlitz-Thouless Realization of 
Superconductivity in the $SU(2) \otimes U_S(1)$ model}

This section is mainly a review of results that appear
in the literature regarding the model~\cite{[5],[2],fmm}. 
It mainly serves as a comprehensive account 
of the various delicate
issues involved, which play a very crucial r\^ole in the underlying 
physics. It is primarily addressed to the non-experts in the area.  
Only the basic results will be presented; the interested
reader may then find the relevant details in the published 
literature.  

An important issue in the effective gauge theory $SU(2) \otimes U_S(1)$ 
model is the 
existence of a {\it global conserved symmetry}, namely the fermion 
number, which is due 
to the electric charge of the fermions $\Psi$.
The corresponding current is given by 
\be
J_\mu = \sum_{\alpha=1}^{2} {\overline \Psi}^\alpha \gamma _\mu \Psi
_\alpha, \qquad \mu=0,1,2.
\label{fermionnumber}
\ee
This current generates a global $U_E(1)$ symmetry, 
which 
after 
coupling with external electromagnetic fields 
is {\it gauged}. 
In this sense the holon current 
(\ref{fermionnumber}) coincides with the charge transport 
properties of the system.

Some discussion is in order at this point. 
The association of the current $J_\mu$ (\ref{fermionnumber}) 
with an electric current for holons comes about 
due to the similarity of the form of the 
spinors (\ref{2.5}) with the conventional Nambu spinors
appearing in the BCS Hamiltonian for superconductivity. 
Indeed, for the benefit of the reader we remind that 
in such a case the electron operators 
$c_\sigma$ are assemblied, in a particle-hole formalism, 
into two component spinors $\left(c_\uparrow , c_\downarrow^\dagger
\right)$, 
and the resulting Hamiltonian couples in a gauge invariant way to 
an external electromagnetic potential ${\vec A}$  
by making the standard substitution of the momentum 
operator  ${\vec p} \rightarrow {\vec p} - \frac{e}{c}{\vec A}$.
The only difference in our nodal liquid case is that the holon spinors
(\ref{2.5}) come in two `colours' and, as contrasted to 
the generic BCS case, the problem is relativistic due to the restriction
in the nodal excitations. Thus, 
at the level of the continuum effective action 
of the nodal excitations, 
the coupling to electromagnetic potentials is straightforward
by extending the (statistical) gauge covariant 
derivatives in the Dirac kinetic terms (\ref{3.7}) to incorporate
the electromagnetic potential coupling terms
\be
  \int d^3x \frac{e}{c}\sum_{\alpha=1}^{2}{\overline \Psi ^\alpha}
\gamma_\mu A_\mu \Psi_\alpha   
\label{electr}
\ee
where $c$ is the light velocity and 
$e$ is the absolute value of 
the electron charge (for holon excitations the charge is $+e$, for electron
$-e$; in our problem here we concentrate in the holon current). 
The resulting nodal holon electric current 
is given by differentiation 
with respect to $A_\mu$, i.e. by the expression (\ref{fermionnumber}).

Before discussing superconducting properties of the 
system we should remark that, as a result of the constraints
(\ref{3.3}) and the non-diagonal nature of the $\gamma_i,i=1,2$
matrices, 
the {\it spatial} components of the current (\ref{fermionnumber}) 
{\it vanish}, but the temporal component (charge density) is non trivial.
Moreover, given that the constraints (\ref{3.3}) 
do not concern the spinons $z$, this means that there is a phase
of the nodal liquid in which there is {\it no charge transport}, but only 
{\it spin transport}. The non-trivial `spin current' may be thought of as
given by $J_\mu ^{spin} \sim  {\overline z} \partial_\mu z $. 
This situation should be compared with the 
corresponding phase in nodal liquids in the approach of 
ref.~\cite{fisher}, 
where the spinons 
are represented as electrically neutral fermions. 

However, in our model 
there are other possibilities, leading to more complicated
phases, as we shall discuss now. These possibilities 
are realized by implementing the constraints (\ref{3.3}) 
via appropriate lagrange multipliers in the path integral 
over the fermionic variables $\psi^\dagger, \psi$, as we discussed
in section 4 (c.f. (\ref{lagrangemult}),(\ref{deltident2})).  
Expressing the products $\psi_1 \psi_2$ (and their conjugates) 
as spatial components of the current (\ref{fermionnumber}), then, 
one may assume a specific ground state in which the appropriate 
lagrange multipliers 
for the constraint $\psi_1 \psi_2 \sim 0$ (and hermitean conjugate) 
acquire non-zero vacuum expectation values that may be absorbed
by appropriate shifts of the corresponding spatial components of 
the electromagnetic potential ${\vec A}(\vec x, t)$ 
coupled to the current ${\vec J}$. 
As we have already discussed  
in section 4, 
a non-trivial vacuum expectation value for the lagrange multiplier 
$\lambda (x)$ of the 
last of the constraint
(\ref{3.3}) will yield mass terms for the $z$ magnons, whilst
the fermionic part of the constraint may be absorbed by an appropriate
shift of the temporal component of the electromagnetic potential. 
This procedure breaks supersymmetry explicitly but, as we shall argue now,
the existence of supersymmetry before 
coupling to external electromagnetism is crucial in implying 
superconducting properties after coupling to external fields. 

In this framework, the constraints (\ref{3.3}) no longer apply in the 
path integral, and non-vanishing spatial compontents of the electric
current, ${\vec J}$, appear. 
It should be remarked that in such a case the 
mixed colour terms in (\ref{4.7}) do not vanish, and hence the resulting 
effective lagrangian breaks supersymmetry explicitly. 
This was to be expected, anyhow, from the the very presence of external 
(non supersymmetric) 
electromagnetic fields.
However, 
given that the coupling of such contact four fermion interactions
is {\it subcritical} (c.f. (\ref{3.8}),(\ref{3.10})), such interactions
are irrelevant operators in a renormalization-group sense, and hence 
the universality class of the theory (in the infrared) can still be 
determined using the supersymmetric version of the theory
in the absence of any external fields 
(which also satisfies the additional restriction (\ref{cond2})).  
As 
we shall argue below, 
this more general phase is important in that it yields 
unconventional superconductivity for the nodal liquid. 

To this end, we remark that 
in the absence of external electromagnetic potentials, the 
symmetry $U_E(1)$ 
is {\it broken spontaneously} 
in the massive phase for the fermions $\Psi$.
This 
can be readily 
seen by considering the following matrix element (see figure \ref{ktbreak}):

\begin{centering}
\begin{figure}[htb]
\vspace{2cm}
%
\bigphotons
\begin{picture}(30000,5000)(0,0)
\put(20000,0){\circle{100000}}
\drawline\photon[\E\REG](22000,0)[4]
\put(18000,0){\circle*{1000}}
\end{picture}
\vspace{1cm}
\caption{{\it Anomalous one-loop Feynman matrix element,
leading to a Kosterlitz-Thouless-like breaking of the 
electromagnetic $U_{em}(1)$ symmetry, and thus 
superconductivity, once a fermion 
mass gap opens up. The wavy line represents the $SU(2)$ 
gauge boson $B_\mu^3$,
which remains massless, while the blob denotes an insertion 
of the fermion-number
current  $J_\mu={\overline \Psi}\gamma_\mu \Psi$.
Continuous lines represent fermions.}}
\label{ktbreak}
\end{figure}
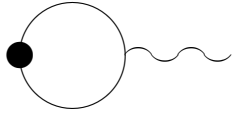
\end{centering}

\be
    {\cal S}^a = <B^a_\mu|J_\nu|0>,~a=1,2,3~; \qquad J_\mu ={\overline
\Psi}\gamma _\mu \Psi 
\label{matrix}
\ee
As a result 
of the colour group structure only the massless $B^3_\mu $ 
gauge boson of the $SU(2)$ group, corresponding to the $\sigma _3$
generator in two-component notation, contributes to the graph. 
The result is~\cite{[2],RK}:
\be
    {\cal S} = <B^3_\mu|J_\nu|0>=({\rm sgn}{M})\epsilon_{\mu\nu\rho}
\frac{p_\rho}{\sqrt{p_0}} 
\label{matrix2}
\ee
where $M$ is the parity-conserving fermion mass 
(or the holon condensate in the context of the 
doped antiferromagnet). 
In our case this mass is generated {\it dynamically} 
by means of the $U_S(1)$ interactions, as we discussed above, 
provided its coupling constant is sufficiently 
strong. 
The result (\ref{matrix2}) is {\it exact} in perturbation theory,
in the sense that the only modifications coming from higher loops 
would be a multpilicative factor $\frac{1}{1-\Pi (p)}$ 
on the right hand side, 
with $\Pi (p)$ the $B_\mu^3$-gauge-boson vacuum 
polarisation function~\cite{RK}.

As discussed in \cite{[2],RK}, 
the $B^3_\mu$ colour component
plays the r\^ole of the {\it Goldstone boson}
of the spontaneously broken fermion-number symmetry.
If this  symmetry is exact, then the gauge boson $B_\mu^3$ 
remains {\it massless}. 
This  
is crucial for the superconducting properties~\cite{[2]}, given that 
this leads to the appearance of a {\it massless pole} in the 
electric-current two-point correlators, the relevant graph being depicted in
figure \ref{pole}. This is the standard Landau criterion for 
superconductivity.

\begin{centering} 
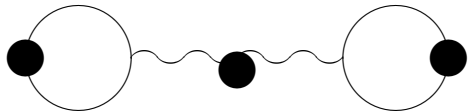
\begin{figure}[h]
\bigphotons
\begin{picture}(5000,5000)(0,0)
\put(15000,0){\circle{100000}}
\drawline\photon[\E\REG](17000,0)[8]
\put(13000,0){\circle*{1400}}
\put(\pmidx,-450){\circle*{1400}}
\global\advance\photonbackx by 2000
\put(\photonbackx,0){\circle{100000}}
\global\advance\photonbackx by 2000
\put(\photonbackx,0){\circle*{1400}}
\end{picture}
\vspace{3cm}
\caption{{\it The lowest-order
contribution to the electric current-current correlator $\langle
0|J_{\mu}(p)J_{\nu}(-p)| 0\rangle$. The
blob in the propagator for the gauge boson $B_\mu^3$ 
indicates fermion loop (resummed) corrections. 
The blob in 
each fermion loop indicates an insertion of the current $J_{\mu}$.}}
\label{pole} 
\end{figure}
\end{centering}

It can be shown~\cite{[2]} that 
in the massive-fermion (broken $SU(2)$) 
phase, the effective low-energy theory 
obtained after integrating out the massive fermionic degrees of freedom
assumes the standard London action for superconductivity, the massless
excitation $\phi$ being defined to be the {\it dual} of $B_\mu^3$:
\be
\partial_\mu \phi \equiv \epsilon_{\mu\nu\rho}\partial_\nu B_\rho^3
\label{dual}
\ee
All the standard properties of superconductivity, Meissner effect (strongly
type II~\cite{[2]}), flux quantization and infinite conductivity, follow 
then in a standard way after coupling to external 
elelctromagnetic potentials, 
provided the excitation $\phi$ (and, hence, $B_\mu^3$)
is exactly massless. 

However, it is known~\cite{[2],[3],[5]} 
that 
superconductivity is of 
a Kosterlitz-Thouless (KT) 
type superconductivity,
not characterized by a local order parameter. 
Let us briefly review the arguments leading to this~\cite{[2]}. 
The neutral parity-invariant condensate 
$<{\overline \Psi}_1 \Psi _1 - {\overline \Psi}_2 \Psi _2 > $,
generated by the strong $U_S(1)$ interaction, 
is {\it invariant} under the $U(1) \otimes U_E(1)$, as a result 
of the $\tau_3$ coupling of $B_\mu^3$ in the action, and hence 
does not constitute an order parameter. This is a characteristic
feature of our gauge interactions.
Putative charge $2e$ or $-2e$  order parameters, 
like the pairing interactions 
among opposite spins in the statistical model of \cite{[2],[5]}, 
e.g. $<\Psi _1 \Psi _2>$, $<{\overline \Psi}_1 
{\overline \Psi}_2>$~\footnote{In four-cmponent notation, such
fermionic bilinears correspond to 
$<\Psi \gamma _5 \Psi >$, $<{\overline \Psi} \gamma _5 {\overline \Psi}>$,
considered in \cite{[2]}.}
will vanish at any finite 
temperature, 
in the sense that strong phase fluctuations will destroy the 
vacuum expectation values of the respective operators, 
due to the Mermin-Wagner theorem. 
Even at zero temperatures, however, such vevs yield zero result 
to any order in perturbation theory trivially, due to the fact that 
in the context of the effective $B_\mu^3$ gauge theory of the broken 
$SU(2)$ phase, the gauge interactions preserve `flavour'. 
For a more detailed discussion  
on the symmetry breaking patterns 
of $(2+1)$-dimensional gauge theories,
and the proper definition of order parameter fields, 
we refer the reader to the 
literature~\cite{RK,[2]}. 
Thus, 
from the above analysis it becomes clear that 
gap formation, pairing and 
superconductivity can occur in the above model without implying any 
phase coherence. 

\subsection{Instantons and the fate of Superconductivity in the $SU(2)
\otimes U_S(1)$ model}

An important feature of the non-Abelian model 
is that, due to the  non-Abelian 
symmetry breaking pattern 
$SU(2) \rightarrow U(1)$, 
the abelian subgroup $U(1) \in SU(2)$,
generated by the 
$\sigma^3$ Pauli generator of $SU(2)$,
is {\it compact}, and 
may contain {\it instantons}~\cite{ahw}, which in three space-time
dimensions 
are like monopoles, and are known to be responsible for giving 
a {\it small} but {\it non-zero mass} to the gauge boson $B_\mu^3$, 
\be 
        m_{B^3} \sim e^{-\frac{1}{2}S_0} 
\label{instmass} 
\ee
where $S_0$ is the one-instanton action, in a dilute gas approximation.
Its dependence on the coupling constant $g_2 \equiv g_{SU(2)}$ 
is well known~\cite{ahw}:
\be
     S_0 \sim \frac{{\rm const}}{g_2^2} 
\label{su2inst}
\ee
For weak coupling $g_2$ the induced gauge-boson 
mass can be very small. However, even 
such a small mass is sufficient to destroy superconductivity,  
since in that case there is no massless pole 
in the electric current-current correlator.
In \cite{fmm} 
a breakdown of 
superconductivity due to instanton effects 
has been interpreted as 
implying a ``pseudogap'' phase: a phase in which 
there is dynamical generation of a mass gap for the 
nodal holons, which, however, is not characterized 
by 
superconducting properties.

The presence of {\it massless} fermions, with 
zero modes around the instanton
configuration, 
is known~\cite{ahw} to suppress the instanton effects on the mass 
of the photon, and under certain circumstances, to be 
specified below, the Abelian-gauge boson may remain exactly massless
{\it even in the 
presence of non-perturbative effects}, thus leading to superconductivity,
in the context of our model. 
This may happen~\cite{ahw} 
if there are extra global symmetries
in the theory, whose currents 
connect the vacuum to the 
one -gauge-boson state, and thus they break spontaneously. 
This is precisely the case of the fermion number 
symmetry 
considered above~\cite{ahw,RK}. In such a case, 
the massless gauge boson is the Goldstone boson of the 
(non-perturbatively) spontaneously broken symmetry. 
However, in our $SU(2) \otimes U_S(1)$ model~\cite{[5],fmm}, 
as a result of the 
(strong) 
$U_S(1)$ interaction, a mass for the fermions is generated,
and hence there is no issue of fermion zero modes in this case.
The analysis of the low energy effective theory 
presented in \cite{[5],fmm} 
is based on a Wilsonian treatment, where massive degrees of 
freedom are integrated out in the path integral. This includes the 
gapful fermions and the massive $SU(2)$ gauge bosons. 
The resulting effective theory, then, which encodes the dynamics 
of the gapped 
phase, is a pure gauge theory 
$U(1) \in SU(2)$, and the instanton contributions to the mass of $B_\mu^3$
are present, given by (\ref{instmass}), in the one-instanton case. 
Thus, it seems that, generically, in the context of the 
$SU(2) \otimes U_S(1)$ of ref. \cite{[5]}, the nodal gap is actually 
a pseudogap.  

\subsection{Instantons and Supersymmetry}

We now remark that 
Supersymmetry is known~\cite{ahw} to suppress instanton contributions.
For instance, in certain $N=1$ supersymmetric models with massless fermions,
considered in ref. \cite{ahw}  
the instanton-induced mass of the Abelian gauge boson is given by:
\be 
         m_{gauge~boson} \sim e^{-S_0} 
\label{instmass2}
\ee
which is suppressed compared to the non-supersymmetric case
(\ref{instmass}).

$N=2$ supersymmetric theories in three space-time 
dimensions
constitute additional examples of theories where the abelian gauge boson 
remains exactly massless, in the presence of instantons~\cite{ahw,dorey}. 
Such theories have complex representation for fermions, and hence 
are characterized by extra global symmetries (like fermion number). 
In view of our discussion above, such models will then lead to 
Kosterlitz-Thouless superconductivity 
upon gauging the fermion number symmetry.

\begin{figure}[htb]
\epsfxsize=4in
\bigskip
\centerline{\epsffile{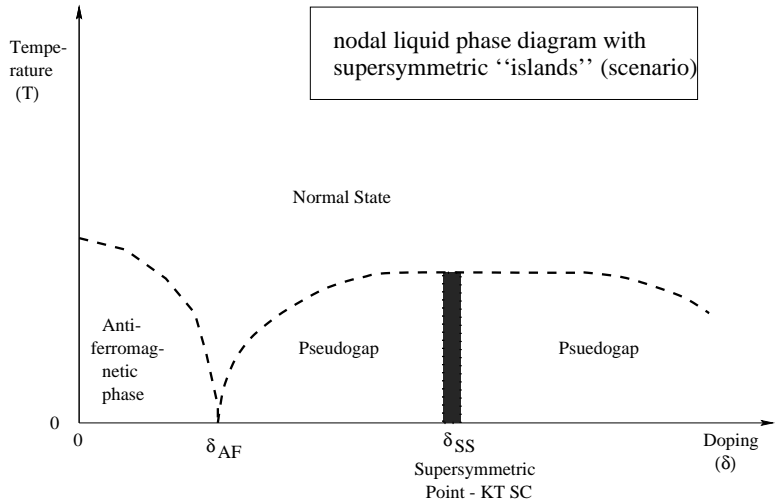}}
\vspace{0.2in}
\caption{\it A possible scenario for the 
temperature-doping phase diagram 
of a charged, relativistic, nodal liquid 
in the context of spin-charge separation.
At certain doping concentrations ($\delta_{SS}$) 
there are
dynamical supersymmetries among the spinon and holon 
degrees of freedom, responsible for yielding   
thin ``stripes'' in the phase diagram (shaded region)
characterized by 
Kosterlitz-Thouless (KT) superconductivity
without a local order parameter. The diagram 
is conjectural at present. It pertains strictly
to the nodal liquid excitations about the 
d-wave nodes of a superconducting gap, and 
hence,  
should not 
be confused with the phase diagram of the entire 
(high-temperature) superconductor.} 
\label{nodalfigure} 
\end{figure}

In this respect,  
the supersymmetric points (\ref{4.1b}),(\ref{cond2}) 
for which such instanton effects are 
argued~\cite{fmm} to be strongly suppressed in favour 
of KT superconductivity, as reviewed above, 
would constitute 
``superconducting stripes'' in the temperature-doping 
phase diagram 
of the nodal liquid 
(see fig. \ref{nodalfigure})~\footnote{It should be stressed
that the term ``stripe'' here is meant to denote 
a certain region of the temperature-doping phase 
diagram 
of the nodal liquid and should not be confused 
with the stripe structures in real space which 
characterizes the cuprates at special doping concentrations.}. 
Theoretically, the stripes should have zero thickness,
given that they occur for specific doping concentrations
(\ref{4.1b}),(\ref{cond2}). However, in practice, there may be uncertainties
(due to doping dependences) in the precise value 
for the parameter $\Delta _z$ entering (\ref{4.1b}),(\ref{cond2})  
which might be 
responsible for giving the superconducting stripe a
certain (small) thickness. 
A detailed analysis of such important issues
is still pending. It is hoped that due to supersymmetry 
one should be able to discuss some exact analytic 
results
at least for zero temperatures.

We also remark that 
in supersymmetric theories of the type considered 
here and in ref. \cite{[8]},
it is known~\cite{ahw} that  
supersymmetry cannot be broken, due to the fact that the Witten 
index $(-1)^F$, where $F$ is the fermion number, is always non zero. 
Thus, in supersymmetric theories the presence of instantons 
should give a small mass, if at all, in {\it both} the gauge boson and the 
associated gaugino, 
However, 
in three dimensional 
supersymmetric gauge theories 
it is possible that 
supesymmetry is broken by having the system in a `false' vacuum,
where the gauge boson remains massless, even in the presence of 
non perturbative configurations, while the gaugino acquires a 
small mass, through non perturbative effects. 
The life time, however, of this false vacuum is very long~\cite{ahw}, 
and hence superconductivity can occur, in the sense that 
the system will remain in that false vacuum 
for a very long period of time, longer than any other time scale 
in the problem.

\subsection{Some Comments on Supersymmetry Breaking at finite temperatures} 

So far, our discussion was restricted to zero temperature. 
At {\it any} finite temperature, no matter how small, 
supersymmetry 
is explicitly broken, 
and thus the supersymmetric points should be viewed as 
{\it quantum critical points}. However, the breaking of supersymmetry
is associated with different boundary conditions 
between fermionic and bosonic degrees of freedom, and,
although the vacuum energy is no longer zero,  
however 
a detailed analysis should be made in order to 
determine whether the equality of mass gaps between the nodal 
spinons and holons at the supersymmetric points 
is lifted by temperature-dependent corrections.
In the context of a supersymmetric theory this issue can be tackled
by means of ``thermal superspace'' methods, which have been developed
recently in the context of particle-physics models~\cite{derendinger}. 
The generic result of such analyses seems 
to be that the mass degeneracy among the 
superpartners is lifted at the level of the 
mass of the various thermal modes,
the corresponding lifting being proportional to the temperature.
The thermal superspace method  
can be applied to the present model
as well, however  
this falls beyond the scope of 
the present article and is thereby left for a future work. 

Moreover,  
as the crude analysis of \cite{fmijmpb} indicates, 
the nodal gaps would disappear at temperatures which are much lower than the
critical temperature of the (bulk) $d$-wave superconducting gap. 
For instance, for a typical set of the parameters of the $t-j$ model
used in \cite{fmijmpb},
the nodal critical temperature is of order of a few $mK$, which is much 
smaller than the $100$ K bulk critical temperature    
of the high temperature superconductors. 
The application of an external magnetic field in the strongly type II
high-temperautre superconducting oxides,
which is another source for explicit breaking of the 
potential supersymmetry, 
enhances the critical 
temperature~\cite{fmijmpb}
up to $30$ K, thereby providing a potential explanation 
for the 
recent findings of \cite{ong}, according to which 
plateaux in the thermal conductivity as a function of the 
external magnetic field indicate the opening of a gap 
at the $d$-wave nodes.

We now remark that, if such situations with broken supersymmetry are viewed 
as cases of perturbed supersymmetric points, then one might 
hope of obtaining non-perturbative information on the 
phase structure of the liquid 
of nodal excitations in spin-charge 
separating scenaria of (gauge) 
high-temperature superconductors. This 
may also prove useful for a complete physical 
understanidng of the 
entire phenomenon, including excitations 
away from the nodes.

\section{Conclusions} 

{}From the above discussion it is clear 
that supersymmetry 
can be achieved in 
the effective continuum field theories of 
doped antiferromagnetic systems 
exhibiting spin-charge separation 
only for {\it particular doping concentrations} 
(cf. (\ref{4.2b}),(\ref{cond2})). 
One's hope is that the ancestor lattice model
will lie in the same universality class (in the infrared) 
as the continuum model, 
in the sense that it differs from it only by the action of 
renormalization-group irrelevant operators. This remains to be checked
by explicit lattice calculations. We should note at this stage
that this is a very difficult problem; in the context
of four-dimensional particle-physics models it is still 
unresolved~\cite{lattice}. 
However, in view of the apparent simpler form of the three-dimensional 
lattice models at hand, one may hope that these 
models are easier to handle.

By varying the doping concentration in the 
sample, one goes away from the supersymmetric point and 
breaks supersymmetry explicitly at zero temperatures. At finite 
temperatures,
or under 
the influence of external 
electromagnetic fields at the nodes of the d-wave gap,  
supersymmetry 
will also be broken explicitly. 
Therefore, realistic systems observed in nature will 
be characterized by explicitly broken supersymmetries
even close to zero temperatures. 
However there is value in deriving such supersymmetric results
in that at such points 
in the parameter space of the condensed-matter system
it is possible to obtain analytically 
some exact results on the phase structure of the theory. 
Supersymmetry may allow for a study of the 
quantum fluctuations about some 
exact ground states of the spin-charge separated
systems in a controlled way. 
Then one may consider perturbing around such exact solutions 
to get useful information about the non-supersymmetric models.

We have argued 
that such special 
points will yield 
new phases for the 
liquid of excitations about 
nodal points of the d-wave superconducting gaps,
which include a phase 
in which there is only spin transport but not
electric current transport, as well as 
a phase in which there are 
Kosterlitz-Thouless type superconducting ``islands''
in a temperature doping phase diagram 
of the nodal liquid, upon the dynamical 
generation of holon-spinon mass gaps (of equal size). 
The latter property is 
due to special properties of the supersymmetry, 
associated
with the suppression of non-perturbative effects of the 
(compact) gauge fields entering the spin-charge 
separation ansatz (\ref{1.5}).  
This, of course, needs to be checked 
explicitly by carrying out the appropriate instanton 
calculations in the spirit 
of the non-perturbative modern framework of \cite{[15]}. 
At present, such non-perturbative effects 
can only be checked explicitly in three dimensions 
for highly extended supersymmetric models~\cite{dorey}.
It is, however, possible that
some exact results could be obtained at least for 
the $N=2$ 
supersymmetric 
models which may have some relevance for the
effective theory of the nodal liquid at the supersymmetric 
points~\cite{[8]}. Then, one may get 
some useful
information for the 
$N=1$ models studied here by viewing them as 
supersymmetry-breaking perturbations  
of the $N=2$ models.
Such issues remain for future investigations,
but we hope that the speculations 
made in the present 
work provide 
sufficient motivation to carry out 
research along these directions.

\par

\par\noindent
\section*{Acknowledgements}\par\noindent

It is a pleasure to acknowledge
informative discussions with I.J.R.~Aitchison, J.~Betouras, 
K.~Farakos, J.H.~Jefferson and G.~Koutsoumbas.  
A preliminary account of this work was presented
by N.E.M. at the Workshop of {\it Common Trends 
in Particle and Condensed Matter Physics Corfu 1999},
Corfu (Greece), 25-28 September 1999 (cond-mat/9909310).
We thank the organizers and participants of this meeting 
for their interest in our work. 
The work of N.E.M. is partially supported by
a P.P.A.R.C. (U.K.) Advanced Fellowship.

\newpage

\par\noindent
\section*{Appendix A} 
\paragraph{}
\noindent{\bf Renormalization aspects of 
Four-Fermi Theories in fewer than four space-time dimensions}
\paragraph{}  
In this appendix we shall review briefly the renormalization-group  
approach to relativistic theories with four-fermion interactions
in fewer than four space-time dimensions.
Below we shall outline only the basic results. 
For further details we refer the interested reader to ref. \cite{[9]}.

We shall use as our pilot theory a three-dimensional 
model, with four-component spinors, containing  
Gross-Neveu interactions. The lagrangian is given by: 
\be {\cal L} = {\overline \psi} i \not D \psi + {g^2 \over 2N}\left(
{\overline \psi}_i \psi_i \right)^2 
\label{A.1}\ee
where $i=1, 2, \dots N$, $N$ 
is a fermion species (`flavour') number, which is assumed large, and 
repeated indices $i$ denote summation.  

Linearizing, by means of a Hubbard-Stratonovich scalar ($\sigma$) field
the four-fermion interactions yields:
\be
{\cal L} = {\overline \psi} i \not D \psi - {1 \over 2g^2} \sigma^2 + 
{1 \over \sqrt{N}} \sigma {\overline \psi}_i \psi_i \label{A.2}\ee  

By naive power counting the four-fermion terms
are irrelevant non-renormalizable operators. 
However, 
the basic observation~\cite{[9]} 
was that in the large $N$ limit
the ultraviolet behaviour 
of the fermion propagator is 
softened in such a way that the scaling dimension of the 
composite 
operator ${\overline \psi}\psi$ changes from its naive dimension,
so that the four-fermion interactions become renormalizable.

This can be seen as follows: 
from (\ref{A.2}) we observe that the tree level scalar propagator 
is given by 
\be
G^{(0)} (p) = g^2 \label{A.3a}
\ee
Consider now the one-loop fermion-vacuum polarization graph.
Assume that a fermion mass $m$ is generated dynamically. $m$ can be
determined
self-consistently by a Schwinger-Dyson approach~\cite{[9]}. 
For our purposes the details of the derivation will be omitted.  
The result for the one-loop vacuum polarization graph is~\cite{[9]} 
\bea
&~&\Pi (p) = -{\rm Tr}\int { d^3k \over 8\pi^3} {1 \over \not k + m }
{1 \over \not k - \not p  + m } = \nn \\
&~&{1 \over \pi} \left[ {m \over 2} 
+ {p^2 - 4 m^2 \over 4 \sqrt{p^2}}{\rm arcsin}
\left(\sqrt{{p^2 \over p^2 + 4m^2}}\right)\right]\equiv -2F(p)\label{A.3b}
\eea

In the large $N$ limit the loop graphs can be resummed. 
To leading order  in $1/N$ expansion
the dressed propagators can be expresssed as
\be
G(p) = {1 \over g^{-2} + 2 F(p)}\label{A.4}\ee

From this epxression it is obvious that 
in the ultraviolat 
limit $p \rightarrow \Lambda >> m$, 
where $\Lambda$ an ultraviolet (high-energy) cut-off,
the behaviour of the propagator $G(p)$ is such that 
$G(p) \sim {1 \over p}$, which implies that the scaling (mass) 
dimension 
of the field $\sigma$ in the ultraviolet regime is $\left[ \sigma
\right]_{UV} = 1$.
From the action (\ref{A.2}) 
it is obvious that the field $\sigma $ is equivalent to 
the composite field ${\overline \psi} \psi $, as far as scaling dimension 
is concerned. 
This implies then that 
the mass dimension of the four-fermion Gross-Neveu operator 
is
\be
\left[\left( {\overline \psi} \psi \right)^2\right]_{UV} = 2~.\label{A.5}
\ee

This guarantees the renormalizability of the theory, since the pertinent 
operators became relevant in a renormalization-group sense:  
if one computes the effective quantum corrections 
to the four-fermion scattering amplitude
in the resummed ${1 \over N}$ approximation, then 
the result for the renormalized coupling is given by $G(p)$ in (\ref{A.4})
which for $p >> m$ scales like 
\be
g_{R}^2 = { g^2 \over 1 - {p \over 8}g^2}
\ee
showing that the effective interaction grows strong for high momenta
for real $g$ (attractive four fermion interactions in our notation). 
Notice that for the four-fermion theory
the renormalizability concerns the 
ultraviolet (high-energy) regime. 
There 
is an UV stable non-trivial fixed point in the theory
and the associated critical exponents can be computed 
within the ${1 \over N}$ 
expansion up to order ${1 \over N^2}$~\cite{[9]}. 
Such computations have also been compared successfully with 
corresponding results from 
lattice simulations.

A Schwinger-Dyson analysis for mass generation~\cite{[9]} 
leads, in the large $N$ limit, to the following gap equation:
\be
 t~m = 4g^2 \int {d^3q \over (2\pi)^3} {m^3 \over q^2 ( q^2 + m^2)}
\label{A.6}
\ee  
with $m$ the dynamically generated mass, and $t={g^2 - g_c^2 \over g_c^2}$,
and the critical coupling $g_c$ is defined through
$1=4g_c^2\int {d^3q \over 8 \pi^3} {1 \over q^2}$. 

Thus, mass generation occurs only for {\it positive } four-fermion couplings
(attractive) Gross-Neveu interactions, which are stronger than a 
given critical value. In a renormalization group sense the repulsive 
interactions are {\it irrelevant}, becoming weaker and weaker 
as one lowers the momenta.

Since the supersymmetric version of the $CP$ $\sigma$-model, of interest
to us here, contains -
as we discuss in Appendix B -  
{\it both} Gross-Neveu and Thirring $({\overline \psi} \gamma _{\mu} \psi
)^2$ 
interactions in it component form~\cite{[8]},   
we turn next our attention to 
a brief review of a 
renormalization group study of such mixed models. 

Such models have been discussed in the literature~\cite{[9]},
with the conclusion that it is mainly the Gross-Neveu interactions 
which determine the critical behaviour, in a large $N$ framework.
Let us review the situation briefly. 
The lagrangian is given by: 
\be
{\cal L} = {\overline \psi} i \not D \psi + {g^2 \over 2N}\left( {\overline
\psi}_i \psi_i \right)^2 + {h^2 \over 2N}\left( {\overline \psi}_i 
\gamma _\mu \psi_i \right)^2 
\label{A.7}
\ee
where $i=1, 2, \dots N$, $N$ 
is a fermion species (`flavour') number, which is assumed large, 
$h^2$ 
is the coupling of the Thirring interactions, 
and, as before,  
repeated indices denote sunmmation.

The Thirring interactions become renormalizable in the UV, just as
the Gross Neveu ones, which can be proven in a similar way to the 
Gross Neveu interactions above, i.e. by linearizing the 
Thirring interaction by a Habbard-Stratonovich {\it vector} field 
$A_\mu$. The vector interactions are viewed as {\it gauge fixed}
interactions
with a bare propagator~\cite{[9]} 
\be
\Delta_{\mu\nu} (p) = h^2 \left(\delta_{\mu\nu} - p_\mu p_\nu/p^2 \right) 
+ {\rm gauge-fixing~terms} \label{A.8}\ee

The dressed (in $1/N$ expansion ) vector propagator 
is modified 
\be
\Delta_{\mu\nu} (p) = {1 \over h^{-2} + F(p)} 
\left(\delta_{\mu\nu} - p_\mu p_\nu/p^2 \right) 
+ {\rm gauge-fixing~terms} 
\label{A.9}
\ee 
where $F(p)$ has been defined in (\ref{A.3b}).

In the ultraviolet regime $p \rightarrow \Lambda$ the scaling mass dimension

of the vector field $A_\mu \sim {\overline \psi} \gamma _\mu \psi$ 
is again one, leading to a renormalizable Thirring interaction 
in the ultraviolet. 

A detailed analysis~\cite{[9]} 
of the critical behaviour in this combined Gross-Neveu 
and Thirring model shows that the critical behaviour is driven by the 
UV fixed point of the Gross-Neveu terms. 
Moreover, repulsive Thirring terms cannot lead to dynamical mass generation,
and thus 
do not affect the critical (fixed point) behaviour of the theory. 

This analysis implies 
that, up to irrelevant operators in a renormalization group sense,
from the various four-fermion contact interactions in our effective theory, 
the Gross-Neveu type interactions appearing in (\ref{3.8}) 
are the only ones that could affect the universality class of the model,
leading to a non trivial Ultraviolet stable fixed point.
However, 
in the context of the planar 
condensed matter systems with relativistic fermions
we are discussing here, 
the Gross-Neveu four-fermion contact interactions are 
sub-critical (c.f. (\ref{3.10})), and hence 
irrelevant operators
in the infrared (low energy) limit. 
In our systems it is the 
gauge-field-holon interactions that grow strong for low momenta and are
thus relevant in a renormalization group sense. This point 
has been discussed in detail in \cite{[21]} 
where we refer the interested reader. In fact, such 
interactions have been argued to 
be responsible for a non-fermi liquid behaviour of the pertinent 
relativistic liquids. 

\section*{Appendix B} 
\paragraph{} 
\noindent{\bf N=1 Supersymmetric $CP^1$ $\sigma$-models in (2+1)-dimensions}

\paragraph{} 

In this Appendix we shall be interested
in discussing briefly the formalsim underlying supersymmetrization 
of 
a $CP^1$ model coupled to Dirac fermions:
\be
{\cal L}_2 = g_1^2|(\partial _\mu 
-  a_\mu  )z|^2  
+i{\overline \Psi}D_\mu\gamma_\mu\Psi
\label{B.1}
\ee
where now $D_\mu = \partial_\mu -ia_\mu $, 
$g_1^2$ has dimensions of mass, 
$a_\mu$ is the $U_S(1)$ (`fractional statistics') field. 
For simplicity we 
consider as a gauge interaction that of a standard $U_S(1)$ Abelian 
gauge theory. 

We consider 
the standard 
$CP^1$ constraint: 
\be
\sum _{\alpha=1}^{2} |z_\alpha|^2 = 1. 
\label{B.8}
\ee
As we shall discuss immediately below, this form of the 
constraint can be supersymmetrized. 

We now proceed to the supersymmetrization of the model 
(\ref{B.1}) 
with the constraint (\ref{B.8}).
Below we shall outline only the main results. For details
we refer the reader to ref.  
\cite{[8],[11]} and references therein.
The main idea behind such a supersymmetrization is to view 
the magnons $z$ as {\it supersymmetric partners} of the holons $\Psi$.

The basic ``matter" multiplet of N=1 supersymmetry in three 
space-time dimensions,
can be written in terms of a scalar superfield as 
\be
\Phi = \phi + \bar{\theta} \chi +(1/2) \bar{\theta} \theta F
\label{B.9}
\ee
which contains a real scalar field, $\phi$, a Majorana spinor
$\chi$ and a real auxiliary field $F$.
We consider complex superfields
\be
Z = (1/\sqrt2)(\Phi_1 + i\Phi_2) = 
z + \bar{\theta} \Psi + (1/2) \bar{\theta} \theta F
\label{B.10}
\ee
which contain a complex scalar, $z=(1/\sqrt2)(\phi_1 + i\phi_2)$,
a Dirac spinor, $\Psi=(1/\sqrt2)(\chi_1 + i\chi_2)$, and a
complex auxiliary field, $F=(1/\sqrt2)(F_1 + iF_2)$.
The supersymmetry transformations read,
\bea 
&~&
\delta_S z = \bar{\xi} \Psi \nn \\
&~&\delta_S \Psi = -i \gamma^{\mu} \xi \partial_{\mu} z + \xi F \nn \\
&~& \delta_S F = -i\bar{\xi} \not{\partial} \Psi
\label{B.11} 
\eea
and the supersymmetric invariant lagrangian is given by the highest
component ($\bar{\theta} \theta$) of the superfield
$\bar{D} Z^* DZ$, 
where
\be
D_{\alpha} = { \partial \over \partial {\overline \theta}_{\alpha}} -
i (\not{\partial} \theta)_\alpha \label{B.12}\ee
is the supersymmetry covariant derivative.

The gauge field is incorporated in a real spinor superfield which,
in the Wess-Zumino gauge, takes the form
\be
V_{\alpha} = i (\not{a} \theta)_{\alpha} + {1 \over 2} \bar{\theta}
\theta \eta_{\alpha}
\label{B.13} 
\ee
where $\eta_\alpha$ 
is the supersymmetric partner of the gauge field (gaugino).

The supersymmetric  gauge invariant lagrangian for the matter fields
which in terms of superfields is the highest component of the superfield
\be
{\overline {\cal D}} Z^* {\cal D} Z
\label{B.14}
\ee
with
\be
{\cal D}_{\alpha} = D_{\alpha} - iV_{\alpha}
\label{B.15}
\ee
In terms of component fields the lagrangian reads:
\be
L = g_1^2 [D_{\mu} \bar{z}^{\alpha}D^{\mu}z^{\alpha} +
i {\overline \Psi} \not{D} \Psi + \bar{F}^{\alpha} F^{\alpha} 
 + 2i({\overline \eta} \Psi^{\alpha} \bar{z}^{\alpha} -
{\overline \Psi}^{\alpha} \eta z^{\alpha})]
\label{B.16} 
\ee
where $D_\mu$ denotes the gauge covariant derivative with respect to the 
$U_S(1)$ field, and for convenience we have rescaled the 
fermion fields $\Psi $ and the auxiliary field $F$ by $g_1$, as compared 
to the non-supersymmetric case, 
in order to facilitate our superfield formalism.

Notice that (\ref{B.16}) contains a supersymmetric 
partner (gaugino) of the statistical gauge field $U_S(1)$. 
This defines the
$N=1$ supersymmetric point of the model, in the sense that the 
gauge interaction $U_S(1)$ `doubles' its 
degrees of freedom as a result of supersymmetry. 
The interactions of the  gaugino 
$\eta$ wth the matter fermion (holon) ${\overline \Psi}$ 
and its partner, the $z$ magnon (spinon),
lead to an 
effective electric-charge violating interactions on the spatial planes,
given that the Majorana gaugino $\eta$ 
is a real field, and as such cannot
carry electric charge (which couples as a phase to a Dirac field). These
terms can be interpreted as the removal or addition of electrons due to
interlayer hopping~\cite{[8]}.

It is important to notice that the constraint (\ref{B.8}) 
admits a $N=1$ supersymmetric formulation, in terms of the superfields 
$Z^\alpha$:
\be
\sum _{\alpha =1}^{2}{\overline Z}^\alpha Z_\alpha = 1  
\label{B.18} 
\ee
which in components yields the constraint (\ref{B.8}) as well 
as~\cite{[8],[11]}: 
\be
{\overline z}_\alpha \Psi _\alpha = 0
\label{cp1constraint2}
\ee
This 
can be solved by means of a `colourless' 
fermion field ${\cal X}$ 
that satisfies (on account of (\ref{B.8})):
\be
   \Psi_\alpha = \epsilon_{\alpha\beta}{\overline z}_\beta {\cal X}~,
    {\cal X}=\epsilon_{\alpha\beta}z_\alpha\Psi_\beta
\label{solution2}
\ee
The auxiliary fields  $F_\alpha$ can be solved by means of their 
equations of motion and the 
 constraint
(\ref{B.18}), or, equivalently, in a 
path integral formalism by implementing the 
constraint via a Lagrange multiplier 
superfield: 
\be \Lambda (x, \theta) = \sigma (x) + {\overline \theta} \delta (x) 
+ {1 \over 2}{\overline \theta} \theta \lambda (x)~. \label{B.18b}\ee
In the second method, by eliminating $F$ one obtains 
$F_\alpha =-\sigma z_\alpha$, for each $\alpha$. If one uses the 
bosonic part of the super-constraint (\ref{B.18}),
$\sum_{\alpha=1}^{2} |z_\alpha|^2=1$,  
one then obtains: 
\be
{\overline F}^\alpha F_\alpha = \sum_{\alpha=1}^{2} {1 \over 4} \left( 
{\overline \Psi}^\alpha \Psi_{\alpha} \right)^2 
\label{B.19}
\ee
We therefore observe that the supersymmetric extension 
of the $CP^1$ $\sigma$ model {\it necessitates} 
the presence of {\it attractive} Gross-Neveu type interactions among the 
Dirac fermions of {\it each sublattice}, 
in addition to the gauge interactions. 
An important point to notice is that the Gross-Neveu terms 
are of the type that would violate parity if dynamical 
generation of fermion mass occured as a result of these interactions. 

For completeness, we also note that 
the presence of Chern-Simons terms, 
which may appear in the effective 
action as a result of parity violation,   
does not add any complciation. 
As discussed in ref. \cite{[11]}, the supersymmetrization 
of these terms leads to a mass for the gaugino, as expected from the 
fact that the Chern-Simons term is a topological gauge boson mass 
term~\cite{[12]}. The result (in components ) is:
\be
S_{CS}^{supersymm} = \int d^3x \kappa \left[  \epsilon _{\mu\nu\rho} 
a_\mu \partial _\nu a_\rho   + {1 \over 4} {\overline \eta} \eta 
\right]\label{B.20}
\ee
where $\eta$ is a Majorana fermion, as we mentioned before, 
and $\kappa$ denotes the coefficient of the Chern-Simons term. 

\paragraph{} 
\noindent{\bf Extension to  $N=2$ Supersymmetric $CP^1$ models} 
\paragraph{}
 
The supersymmetric $N=2$ $CP^M$ $\sigma$-model was constructed 
in ref. \cite{[11]}, by dimensional reduction from 
a four-dimensional $N=1$ supersymmetric  
lagrangian in a Minkowskian space time, which is super gauge and 
$U(M)$ invariant:
\be
iS^{(4)} ={i\over 16}\int d^4x d^2\theta d^2{\overline \theta}
{\overline \Phi} e^V \Phi + {i \over 128g^2} \int d^4x d^2 \theta d^2
{\overline \theta} 
V D{\overline D}V \label{B.21}
\ee   
in a standard four-dimensional superfield notation~\cite{[13]}
with $\theta ^\alpha, {\overline \theta}_{\dot \alpha}$ complex spinors,
and $D_\alpha = {\partial \over \partial \theta^\alpha } - 
i \sigma_{\alpha{\dot \alpha}}^\mu {\overline \theta}^{\dot \alpha}
\partial _\mu, {\overline D}^{\dot \alpha }= 
{\partial \over \partial {\overline \theta}_{\dot \alpha} } + 
i \theta ^\alpha \sigma_{\alpha{\dot \alpha}}^\mu 
\partial _\mu$; the vector superfield $V$ is real, and contains the 
gauge bosons, $A_\mu$, whilst the scalar superfield    
$\Phi$ is chiral $({\overline D}_{\dot \alpha} \Phi =0)$. 
In component form the action (\ref{B.21}) 
reads: 
\bea
&~& S^{(4)} = i \int d^3x dt \left[ {\overline D}_\mu {\bar z} D^\mu z + 
i {\overline \psi} \not D \psi + {\overline F}F + 
\lambda {\bar z}z 
+ {\bar \psi}uz + {\bar u} \psi  {\bar z} \right]
+ \nn \\
&~& i\int d^3x dt \left[{1 \over f^2} \left[ -{1 \over 4} F_{\mu\nu}^2 + 2
i{\overline u}\not\partial u + 2d^2\right]_{irrelevant}\right]
\label{B.22}
\eea
where $D_\mu = \partial_\mu - i A_\mu$, and 
the various component fields can be understood by the field content
of the supersymmetry algebra~\cite{[13]}.
The terms marked irrelevant are, by power counting, 
irrelevant operators in a renormalization groups sense in the low-energy 
regime of the dimensionally-reduced three-dimensional theory,  and hence
they can be safely 
ignored. The dimensional reduction in the Minkowski time $t$ 
leads to a Euclidean three dimensional theory, which is precisely 
a $N=2$ supersymmetric $\sigma$-model$~\left[ 11 \right]$: 
\be
S_E^{(3)} = \int d_E^3x \left[ 
{\overline D}_i{\bar z} D^i z + 
i {\overline \psi} \not D \psi - A_0^2{\bar z}z + A_0{\overline \psi}\psi 
- {\overline F}F + 
\lambda {\bar z}z 
+ {\bar \psi}uz + {\bar u} \psi  {\bar z}\right]\label{B.23}\ee
up to irrelevant operators
of the form ${1 \over f^2} \left[ {1 \over 4} F_{ij}^2 
-{1 \over 2} (\partial _i A_0)^2 
- 2 i{\overline u}\not\partial u -2 \lambda^2 \right]$. 
Notice that the temporal component of 
the (dimensional reduced) four-dimensional gauge potential $A_0$ plays the 
r\^ole of the Lagrange multiplier $\sigma$ field 
in an $N=1$ formulation 
(c.f. (\ref{B.18b}). The important differnce of the $N=2$ formalism,
however, 
is that now the gaugino field ${\bar u}$ is a Dirac spinor. 

A $N=2$ supersymmetric 
version of the Chern-Simons terms also exists~\cite{[11]}.
In component terms is given by
\be
S_{CS}^{N=2~supersymm} = \int d_E^3x \kappa \left[  \epsilon _{\mu\nu\rho} 
a_\mu \partial _\nu a_\rho   + {i \over 4} {\overline u} u
+ \lambda \sigma \right]
\label{B.24}\ee
The mixing $\lambda\sigma$ 
is a feature of the 
$N=2$ formalism and was absent in the $N=1$ case (\ref{B.20}). 

In the context of our statistical spin=charge separating model, 
the presence of a Dirac gaugino allows, in contrast to the 
$N=1$ case,  
for electric charge conservation~\cite{[8]},
given that now the gaugino being a Dirac spinor
is allowed to carry electric charge. Thus the coupling terms
to the matter fermions  
${\overline \Psi} u $ are now electrically neutral.
This would imply suppression of intersublattice or interlayer hopping.

The critical behaviour of the $N=2$ model has been studied in detail
in ref. \cite{[11]}. Due to the extra supersymmetry, 
the $N=2$ case
allows for an {\it exact} 
computation of the pertinent renormalization-group 
$\beta$-functions~\cite{[11],[14]}, and hence the 
critical exponents of the model. 
This has to be contrasted with the situation in the $N=1$ model,
where such exact results are not available.   
Such exact results make the $N=2$
$CP^M$ $\sigma$-model attractive for further studies
along the lines of the fully non-perturbative approach 
to strongly-coupled supersymmetric gauge theories, advocated 
by Seiberg and Witten~\cite{[15],[16]}. 
Then, by viewing the $N=1$ case, of relevance to us in the context
of doped antiferromegnets, as a supersymmetry boken descendant 
of the $N=2$ model, 
one might obtain 
valuable non-perturbative information 
for the phase structure of the theory at the supersymmetric points 
of the parameters of the system.

\end{document}